\documentclass[aps,prd,twocolumn,groupedaddress,showpacs,nofootinbib,amssymb]{revtex4}

\usepackage[T1]{fontenc}
\usepackage[latin1]{inputenc}
\usepackage{graphicx}
\usepackage[english]{babel}
\usepackage{graphicx}
\usepackage{epstopdf}
\usepackage{bm}
\usepackage{amsmath}
\usepackage{amssymb}
\usepackage{amsfonts}
\usepackage{epsfig}
\usepackage{colordvi}
\usepackage{color}
\usepackage{dcolumn}
\usepackage{multirow}
\usepackage{doi}
\usepackage{hyperref}
\hypersetup{
  colorlinks=true,        
  linkcolor=blue,         
  citecolor=cyan,         
}

\graphicspath{{./}}
\newcommand{\be}{\begin{equation}}
\newcommand{\ee}{\end{equation}}
\newcommand{\ba}{\begin{eqnarray}}
\newcommand{\ea}{\end{eqnarray}}

\def\ltsima{$\; \buildrel < \over \sim \;$}
\def\simlt{\lower.5ex\hbox{\ltsima}}
\def\gtsima{$\; \buildrel > \over \sim \;$}
\def\simgt{\lower.5ex\hbox{\gtsima}}

\def\apj{Astrophysical Journal}                

\def\prd{Phys.~Rev.~D}       

\def\dchizeromedian{$-0.05_{-0.1}^{+0.08}$}
\def\dchionemedian{$0.18_{-0.26}^{+0.31}$}
\def\dchitwomedian{$-0.05_{-0.21}^{+0.12}$}
\def\dchithreemedian{$0.11_{-0.18}^{+0.06}$}
\def\dchifourmedian{$-0.6_{-0.82}^{+0.81}$}
\def\dchifivelmedian{$0.27_{-0.4}^{+0.26}$}
\def\dchisixmedian{$-0.38_{-0.72}^{+0.49}$}
\def\dchisixlmedian{$2.66_{-3.53}^{+3.33}$}
\def\dchisevenmedian{$1.48_{-1.73}^{+1.59}$}

\begin{document}

\title{Constraining alternative theories of gravity using GW$150914$ and GW$151226$}

\author{Mariafelicia De Laurentis$^{1,4,5,6}$\footnote{e-mail address:laurentis@th.physik.uni-frankfurt.de}, Oliver Porth$^{1}$\footnote{e-mail address:porth@th.physik.uni-frankfurt.de}, Luke Bovard$^{1}$\footnote{e-mail address:bovard@th.physik.uni-frankfurt.de}, Bobomurat Ahmedov $^{2,3}$\footnote{e-mail address:ahmedov@astrin.uz }, Ahmadjon Abdujabbarov$^{3,4}$\footnote{e-mail address:ahmadjon@astrin.uz}}
\affiliation{$^{1}$Institute for Theoretical Physics, Goethe University, Max-von-Laue-Str. 1, 60438 Frankfurt, Germany}
\affiliation{$^2$National University of Uzbekistan, Tashkent 100174, Uzbekistan}
\affiliation{$^3$Ulugh Beg Astronomical Institute, Astronomicheskaya 33,
Tashkent 100052, Uzbekistan}
\affiliation{$^4$ Tomsk State Pedagogical University, ul. Kievskaya, 60, 634061 Tomsk, Russia}
\affiliation{$^5$ Lab.Theor.Cosmology,Tomsk State University of Control Systems and Radioelectronics (TUSUR), 634050 Tomsk, Russia}
\affiliation{$^{6}$INFN Sezione  di Napoli, Compl. Univ. di
Monte S. Angelo, Edificio G, Via Cinthia, I-80126, Napoli, Italy}


\date{\today}
\begin{abstract}
The recently reported gravitational wave events GW$150914$ and GW$151226$ caused by the mergers of binary black holes~\cite{LIGO16b,LIGO16e,LIGO16f} provide a formidable way to set constraints on alternative metric theories of gravity in the strong field regime. In this paper, we develop an approach where an arbitrary theory of gravity can be parametrised by an effective coupling $G_{eff}$ and an effective gravitational potential $\Phi(r)$. The standard Newtonian limit of General Relativity is recovered as soon as $G_{eff}\rightarrow G_N$ and $\Phi(r)\rightarrow \Phi_{N}$. The upper bound on the graviton mass and the gravitational interaction length, reported by the LIGO-VIRGO collaboration, can be directly recast in terms of  the parameters of the theory which allows an analysis where the gravitational wave frequency modulation sets constraints on the range of possible alternative models of gravity.  Numerical results based on published parameters for the binary black hole mergers are also reported. Comparison of the observed phase of the GW$150914$ and GW$151226$ with the modulated phase in alternative theories of gravity does not give reasonable constraints due the large uncertainties in the estimated parameters for the coalescing black holes. In addition to these general considerations, we obtain limits for the frequency dependence of the $\alpha$ parameter in scalar tensor theories of gravity.
 \end{abstract}
\pacs{04.30.-w; 04.50.+h; 97.60.Lf}

\maketitle
\section{Introduction}
\label{uno}

In September and December $2015$, the LIGO-VIRGO collaboration has reported on the direct detection of gravitational-wave (GW) signals from coalescing binary black hole (BH) systems \cite{LIGO16b,LIGO16c}.  This has opened new opportunities in gravity research and began the era of gravitational wave astronomy.  In particular, this achievement can be considered as the first direct probe of metric theories of gravity in the regime of strong fields and relativistic velocities.
The individual masses of the merging BHs at the beginning of the collision were $29^{+6}_{-4}M_\odot$ and  $36^{+4}_{-5}M_\odot$ for the September signal and $14.2^{+8.3}_{-4.2}M_\odot$ and $7.5^{+2.3}_{-2.3}M_\odot$ for the December signal. Specifically, the GW$150914$ signal was emitted by a rapidly evolving dynamical binary that merged in a fraction of a second with an observed variation of the period ${\dot P}_b$ ranging  from $\sim -0.1$ at $f_{\rm GW} \sim 30\ {\rm Hz}$ to $\sim -1$ at $f_{\rm GW} \sim 132\ {\rm Hz}$.  
The frequency and amplitude of the GW$151226$ signal was observed over $55$ cycles spanning a range in frequency from $35$ to $450$ Hz. 
Using the templates created from numerical relativity, the data is consistent with the merger of two compact objects into a merged black hole with masses of $ \sim 65.3^{+4.1}_{-3.4} M_\odot$ and $ \sim 21.8^{+5.9}_{-1.7}M_\odot$, respectively.
In this process, the energy emitted in the form of gravitational waves (GW) amounts to $3.0^{+0.5}_{-0.4}M_\odot$ and $1.0^{+0.1}_{-0.2}M_\odot$ and the velocity $v$ reached the value $\sim 0.5c$ at the time of the merger.  
In particular, the signal from GW$150914$ exhibits the typical behaviour predicted by the coalescence of compact systems where  inspiral, merger and ring down phases are traversed \cite{LIGO16a}.
The LIGO-VIRGO collaboration has analyzed the three regimes adopting a parametrized analytical family
of inspiral-merger-ringdown waveforms \cite{Blanchet94,Blanchet95b,Arun06,Mishra10,Yunes09,Li12}. The signal is divided in terms of frequency: the early to late inspiral regime  from $\sim 20$\,Hz to $\sim$55\,Hz; the intermediate region from $\sim$ 55\,Hz to $\sim$ 130\,Hz and the merger-ringdown region from $\sim$ 130\,Hz until the end of the waveform.
 The simplest and fastest parameterized waveform model that is currently available~\cite{Husa15} sets bounds on the physical effects based on the inspiral phase only, where a calibrated post-Newtonian (PN) treatment is sufficient.  
For the later phases, phenomenological coefficients fitted to Numerical Relativity (NR) waveforms are used.  
In this paper, we discuss the possibility to set constraints on extended theories of gravity via the modified inspiral phase.  

It is worth noting that the existence of GWs confirms metric theories of gravity, among them General Relativity (GR), but there is ample room for other possibilities (see \cite{LIGO16b} for a detailed discussion).
Any extended theory of gravity can be parameterized by means of a suitable post-Newtonian parametrization where the governing parameter is the effective gravitational coupling constant $G_{eff}$ and the effective gravitational  potential $\Phi (r)$.  Both these quantities are functions of the radial coordinate that influence the phase of the GW signal. 
In other words, the GW waveform could, in principle, single out the range of possible gravitational metric theories that are in agreement with the data.

The paper is organised as follows. In Sec. \ref{due}, we discuss how different theories of gravity can be parametrized by the coupling constant and the gravitational potential.  The main differences of these theories with respect to GR can be reduced to the effective dependence on the radial coordinate. 
It is then straightforward to obtain the corresponding phase modulation and we will exemplarily do so in Sec. \ref{tre} and compare with the observed data. Sec. \ref{quattro} is devoted to the discussion of Shapiro delay that can be modulated according to the parameters of the given theory. Discussion and conclusions are drawn in Sec. \ref{cinque}.
\section{Effective gravitational constant in extended theories of gravity}
\label{due}
Alternative theories of gravity are extensions of GR where higher order curvature invariants and/or additional scalar fields are taken into account in the Hilbert-Einstein gravitational action (see \cite{Capozziello11,Nojiri11,Capozziello10,Olmo11} for a comprehensive review on the subject).
If the gravitational Lagrangian is nonlinear in the Ricci scalar or, more generally, in the curvature invariants, the field equations  become higher than second order in the derivatives; it is for
this reason that such theories are often called {\it higher-order gravitational theories}.
In principle, one can take into account wide classes of higher-order-scalar-tensor theories of gravity in four dimensions \cite{Capozziello11}.

With the emergence of the inflationary paradigm, these theories have gained heightened attention as they can provide solutions to the shortcomings of the standard cosmological model.  These are for example: 
the horizon problem,
the density fluctuation problem ,
the dark matter problem,
the exotic relics problem,
the thermal state problem
the cosmological constant problem
the singularity problem
the timescale problem \cite{Kolb90,Mukhanov81,Guth82,Hawking82,Starobinsky82}.
 
 Furthermore, the presence of
scalar fields is important also in
 multidimensional gravity, such as Kaluza-Klein theories and in
 the effective action of string theory.
In this framework, the strength of gravity, given by the local
value of the gravitational coupling, depends on time and location.
For example, the Brans-Dicke theory, that is the
most used scalar-tensor theory of gravity \cite{Brans61}, includes the
hypothesis suggested by Dirac of the variation of the
gravitational coupling with time \cite{Sciama53}. As a consequence,
scalar-tensor theories  do not satisfy the Strong Equivalence Principle as the variation
of the gravitational constant $G_{eff}$ -- which is, in general,
different from $G_N$, the standard Newton gravitational constant
-- implies that local gravitational physics depends on the scalar
field strength. Theories which present such a feature are called {\it non-minimally coupled theories}.

In these theories,  the gravitational coupling is determined
by the form of the Lagrangian. We can have two physically interesting
situations which could be tested by experiments:

\begin{enumerate}
 \item when $G_{eff}(r)_{r \rightarrow \infty} \rightarrow G_N$, the Newton gravitational
 constant and GR  are recovered.
 \item The possibility that gravitational coupling is not asymptotically constant, {\it i.e.} $G_{eff}$ is
 always
 varying with the epoch  and $\dot{G}_{eff}/G_{eff}\vert_{now}\neq 0$.
\end{enumerate}

The variability of the gravitational coupling can be tested by three classes
of experiments:

\begin{itemize}

 \item Through observations of Solar System dynamics. In fact, several weak-field tests  of GR are based on planetary motion and dynamics of  self-gravitating objects nearby the Sun.  Deviations from classical tests are possible probes for the variation of the gravitational coupling.

\item Through binary pulsar systems.
  In order to obtain information from these systems, it has been necessary
  to extend the post-Newtonian approximation, which can be used only
  in the presence of a weakly gravitationally interacting n-body system,
  to strong gravitationally interacting systems. The estimation of $\dot{G}/G$ is $2 \times 10^{-11}$ per year \cite{Uzan03,Damour96}.

\item Through gravitational lensing observations of distant galaxies \cite{Krauss92}. 

\end{itemize}

Concerning the solar system tests, the most stringent limits are obtained by Lunar Laser Ranging (LLR) combined with accurate ephemeris of the solar system.  
LLR consists of measuring the round-trip travel time of photons that are reflected back to Earth from mirrors located on the Moon;
 the change of round-trip time contains information about the Earth-Moon system.
 The round-trip travel time has been investigated for many years and the best estimates for $\dot{G}/G$ range from $0.4 \times 10^{-11} \rm yr^{-1}$ to $10^{-11} \rm yr^{-1}$ \cite{Dickey94,Uzan03}.
However, none of these tests probes the strong field regime which, up to now, could not be investigated at all.

Besides the variation of the gravitational coupling,
it is well known that a wide class of these theories gives rise to Yukawa-like corrections, $r^{-1} e^{-mr}$ in the gravitational potential \cite{Capozziello10}. Here, the parameter $m$ is an effective
mass related to the additional degrees of freedom in the gravitational action. Specifically,  an additional scalar field is introduced by the corresponding Klein-Gordon equation of the form $\Box\phi-dV(\phi)/d\phi=0$ that has to be added to the standard set of Einstein field equations. In the static case,
the  Klein-Gordon equation reduces to
\begin{equation}
\label{poisson}
(\nabla^2-m^2) \, \phi=0\,,
\end{equation}
where the effective mass $m$ is given by the minimum of the potential $V(\phi)$.  The solution of Eq.(\ref{poisson}) is a Newtonian potential corrected by a  Yukawa-like term that,
as in the Klein-Gordon case,  disappears at infinity allowing to recover the Newtonian limit and Minkowski flat spacetime.

In general,  most alternative gravities have a weak field limit that can be expressed in the form
(see also  \cite{Capozziello12})
\begin{eqnarray}
 \Phi(r) = -\frac{G_N \, M}{r} \, \left[1+\sum_{k=1}^{n} \, \alpha_k \, e^{-r/r_k}\right]=-\frac{G_{eff} \, M}{r}\,, \label{eq:yukawa} \end{eqnarray}
where $G_{N}$ is the value of the gravitational constant as
measured at infinity and $r_k$ is the interaction length of the
$k$-th component of the non-Newtonian corrections (see also \cite{Stelle78,Kenmoku93}).
%
See Ref. \cite{Stelle78,Kenmoku93}, for a general discussion of this last equation containing the non-Newtonian corrections.

Clearly, the standard Newtonian potential is restored as soon as $G_{eff}\rightarrow G_N$, which means $e^{-r/r_k}\rightarrow 0$ at infinity.
The amplitude
$\alpha_k$ of each component is normalized to the standard Newtonian
term and the signs of the $\alpha_k$ coefficients indicate if the
corrections are attractive or repulsive \cite{Will93}.

For the simplicity of the estimations, one can truncate to the first term of the
expansion series in Eq.~(\ref{eq:yukawa})\footnote{This assumption is not applicable in some cases where additional corrections are taken into account.}.

One then obtains a potential for the form
\begin{eqnarray}
 \Phi(r) \;=\; -\frac{G_N \, M}{r} \,\left[1+\alpha_1 \, e^{-r/r_1}\right]\,,
\label{eq:yukawa1} \end{eqnarray}
where the influence of non-Newtonian terms can be parameterized through the constants
$(\alpha_1,\,r_1)$. For asymptotically large distances, where $r\gg r_1$, the
exponential term tends to $0$ and consequently the gravitational coupling tends to
the limiting value
$G_{N}$. In the opposite case when $r\ll r_1$, the exponential term tends to unity, consequently, by
differentiating Eq.~(\ref{eq:yukawa1}) and comparing with the
gravitational force measured in laboratory experiments, one can get
%
%
%
\begin{eqnarray}
 G_{lab} = G_{N} \, \left[1+\alpha_1 \,\left(1-\frac{r}{r_1}\right)\right]
\simeq G_{N} (1+\alpha_1)\,, \nonumber\\ \label{eq:yukawa2}  \end{eqnarray}
where $G_{lab}=6.67\times 10^{-8}$ g$^{-1}$cm$^3$s$^{-2}$
is the standard Newton gravitation constant precisely measured in
Cavendish-like experiments and where $G_{N}$ and $G_{lab}$
are identically the same in the standard gravity. However,
 the inverse square law is asymptotically valid, but the measured
coupling constant is different by a factor $(1+\alpha_1)$.

For self-gravitating systems, any  correction involves a characteristic length that acts at a
certain scale. The range of the characteristic scale $r_k$ corresponds to the Compton's length
\begin{eqnarray}
 r_k \;=\; \frac{\hbar}{m_k \, c} \;
\label{eq:compton} \end{eqnarray}
is identified
through the mass $m_k$ of a pseudo-particle.
Accordingly, in the weak energy limit, fundamental
theories attempting to unify gravity with other forces
introduce extra particles {\it
with mass} which may carry the further degrees of freedom of the gravitational force \cite{Gibbons81}.
%

There have been several attempts to constrain $r_k$ and
$\alpha_k$ (and hence $m_k$) by experiments on scales in the range $1
\,\mbox{cm}<r< 10^8\, \mbox{cm}$, using a variety of independent and different
techniques \cite{Fischbach86,Speake88,Eckhardt88}. The expected masses
for particles which should carry the additional gravitational
force are in the range $10^{-13}\, \mbox{eV}<m_k< 10^{-5}\,
\mbox{eV}$. 
Given these, one can obtain the following estimates for the parameters
\begin{equation} |\alpha_1|\sim 10^{-2}\,,\qquad r_1 \;\sim\; 10^4 - 10^5
\,\mbox{cm} \,.\label{eq:range}
\end{equation}
 Assuming that the dilaton is an ultra-soft boson which
 carries the scalar mode of gravitational field, one obtains
 a length scale of $\sim 10^{22} - 10^{23}$ cm, if the mass range is $m\sim 10^{-27} - 10^{-28}\, \mbox{eV}$.
This length scale is necessary to explain the flat rotation curves of the spiral galaxies.
Furthermore, Very Long Baseline Interferometry observations impose a limit of $\alpha\sim 1.4\times 10^{-2}$ \cite{Eubanks97}. On the other hand, binary-pulsar data places a limit from $10^{-2}$  to $10^{-4}$ on $\alpha$ \cite{Taylor93,Stairs98,Damour91,Wex97}.

However, new limits from GW$150914$, reported in \cite{LIGO16b}, give as an upper limit for the graviton  mass $m_g \leq 10^{-22}\, $eV and $r_g \geq 10^{18}\, $cm for the related Compton length.  We obtain the same limit also for GW$151226$. These experimental numbers  open new interesting perspectives in the present  debate as soon as the above $m_k$ and $r_k$ are interpreted.
Below, we will discuss how $G_{eff}$ and $\Phi(r)$ could be constrained according to the GW$150914$ and GW$151226$ data.  As we will see, such constraints can be interpreted, at fundamental level, as the above effective mass $m_k$ and interaction length $r_k$.

\section{Constraining $G_{eff}$ and $\Phi(r)$ by GW150914 and GW151226 }
\label{tre}

Starting from the above considerations, it is possible to constrain $G_{eff}$ and $\Phi(r)$ by the GW parameters reported for the events GW$150914$ and GW$151226$. Before this, let us review the post-Newtonian approximation required to perform this kind of analysis.
Specifically, let us compute the $3.5$PN approximation that relevant for our analysis \cite{Maggiore08,Weinberg72}. In particular, PN waveform models at the $3.5$PN order are developed {\it e.g.} in \cite{Buonanno09}.

To compare the theoretical waveforms with experimental sensitivities, we write the Fourier transform of the two GW strains $h_{+},h_{\times}$ as
\begin{eqnarray}
h_{+}&=&A e^{i\phi_{+}(f)}\frac{c}{r}\left(\frac{G_{eff}M}{c^3}\right)^{\frac{5}{6}}\frac{1}{f^{\frac{7}{6}}}\left(\frac{1+\cos^2 i}{2}\right)\,,\\
h_{\times}&=&A e^{i\phi_{\times}(f)}\frac{c}{r}\left(\frac{G_{eff}M}{c^3}\right)^{\frac{5}{6}}\frac{1}{f^{\frac{7}{6}}}\cos i\,,
\end{eqnarray}
where $i$ is the inclination angle of the line of sight and the constant $A$ has the value
\begin{eqnarray}
A=\frac{1}{\pi^{\frac{2}{3}}}\left(\frac{5}{24}\right)^{\frac{1}{2}}\,.
\end{eqnarray}
The phase $\phi_{+}$ is given as
\begin{eqnarray}
\phi_{+}(f)=2\pi f\left(t_c+\frac{r}{c}\right)-\varphi_c-\frac{\pi}{4}+\frac{3}{4}\left(\frac{G_{eff}M}{c^3}8\pi f\right)^{-\frac{5}{3}}\,,\nonumber\\
\label{psi}
\end{eqnarray}
where $\varphi_c$ and $t_c$ are the value of the phase and the time at coalescence, respectively. Furthermore the phases of the two strains are directly related, $\phi_{\times}=\phi_{+}+\frac{\pi}{2}$.

An accurate computation of the phase going well beyond the Newtonian approximation, is crucial for discriminating the signal of a coalescing binary from the noise. Therefore one has to give the PN correction to the phase \eqref{psi}. 
In order to exploit the signal present in the detector, and thus detect sources at further distance, an accurate theoretical prediction on the time evolution of the waveform is required.

In order to calculate the PN corrections we write the equation of motion in a more general form
\begin{eqnarray}
\frac{dv^i}{dt}=-\frac{G_{eff}M}{r^2}\left[(1+{\cal A})\frac{x^i}{r}+{\cal B}v^i\right]+{\cal O}\left(\frac{1}{c^8}\right)\,,\nonumber\\
\end{eqnarray}
such that it has a term proportional to the relative separation $x^i$ and a term proportional to the relative velocity $v^i$ in the center of mass frame. Here, the effective gravitational constant is not given by the standard Newton constant, but by $G_{eff}=G_N(1+\alpha)$. 


Explicit expressions for the functions ${\cal A}$ and ${\cal B}$ are extremely long and are given
in Ref.~\cite{Mirshekari13}.
 Here we address the issue of how to obtain constraints from GW150914 and GW151226 data in the framework of the frequency-domain waveform model \cite{LIGO16a}.
We proceed by considering the following relation for the frequency-domain phase
\begin{eqnarray}
&&\phi = 2\pi f t_c-\varphi_c-\frac{\pi}{4}\nonumber\\&&+\frac{3}{128\eta}\left(\pi \frac{M f G_{eff}}{c^3}\right)^{-\frac{5}{3}}\sum_{i=0}^{7}\varphi_i(\Theta)\left(\pi\,\frac{M f G_{eff}}{c^3}\right)^{\frac{i}{3}}\,,\nonumber\\
\end{eqnarray}
where $\varphi_i(\Theta)$
are the PN expansion coefficients that are functions of the intrinsic binary parameters. The information on the spin $\chi_i$ (with $i=1,2$)  is  incorporated via the following relations
\begin{align}
\chi_s &=\frac{(\chi_1 + \chi_2)}{2},\\
\chi_a &= \frac{(\chi_1 - \chi_2)}{2}\,.
\end{align}
that appear in the functions $\varphi_i(\Theta)$.  

The $3.5$PN expansion coefficients are
\begin{align}
  \varphi_0 &= 1,\\
  \varphi_1 &= 0,\\
  \varphi_2 &= \frac{3715}{756}+\frac{55 \eta }{9}, \\
  \varphi_3 &= -16 \pi +\frac{113 \delta  \chi
_a}{3}+\left(\frac{113}{3}-\frac{76 \eta }{3}\right) \chi _s,
\end{align}
\begin{widetext}
 \begin{align}
   \varphi_4 &= \frac{15293365}{508032}+\frac{27145 \eta }{504}+\frac{3085 \eta
^2}{72}+\left(-\frac{405}{8}+200 \eta \right) \chi _a^2-\frac{405}{4} \delta
\chi _a \chi _s+\left(-\frac{405}{8}+\frac{5 \eta }{2}\right) \chi _s^2, \\
\varphi_5 &= \left[ 1 + \log \left( \pi \frac{G_{eff}M f}{c^3} \right) \right] \left[ \frac{38645
\pi }{756}-\frac{65 \pi  \eta }{9}+\delta  \left(-\frac{732985}{2268}-\frac{140
\eta }{9}\right) \chi _a +\left(-\frac{732985}{2268}+\frac{24260  \eta
}{81}+\frac{340 \eta ^2}{9}\right) \chi _s \right], \label{eq:phi5}\\
\begin{split}
\varphi_6 & = \frac{11583231236531}{4694215680}-\frac{6848\gamma_E}
{21}-\frac{640 \pi ^2}{3}+\left(-\frac{15737765635}{3048192}+\frac{2255 \pi
^2}{12}\right) \eta +\frac{76055 \eta ^2}{1728}-\frac{127825 \eta
^3}{1296} \\ & \phantom{=} -\frac{6848}{63} \log \left(64 \pi \frac{G_{eff}M f}{c^3} \right)+\frac{2270}{3}
\pi \delta \chi _a+\left(\frac{2270 \pi }{3}-520 \pi \eta \right) \chi _s, \label{eq:phi6}
\end{split}
\\
\begin{split}
\varphi_7 &= \frac{77096675 \pi }{254016}+\frac{378515 \pi  \eta
}{1512}-\frac{74045 \pi  \eta ^2}{756}+\delta
\left(-\frac{25150083775}{3048192}+\frac{26804935 \eta }{6048}-\frac{1985 \eta
^2}{48}\right) \chi _a \\
& \phantom{=} +\left(-\frac{25150083775}{3048192}+\frac{10566655595 \eta
}{762048}-\frac{1042165 \eta ^2}{3024}+\frac{5345 \eta ^3}{36}\right) \chi _s.
\end{split}
 \end{align}
\end{widetext}
where $\varphi_0,...,\varphi_7$ indicate the $0,...,3.5$PN approximation, respectively and $\gamma_{E} = 0.577$ is the Euler-Mascheroni constant \cite{Yunes09} and we have use the common definitions
\begin{align}
\delta &=\frac{ (m_1 - m_2)}{M},\\
\eta &=\frac{ (m_1m_2)}{M}
\end{align}
where $m_1$, $m_2$ are the masses of the two compact objects.  

\begin{figure*}
\includegraphics[scale=0.38]{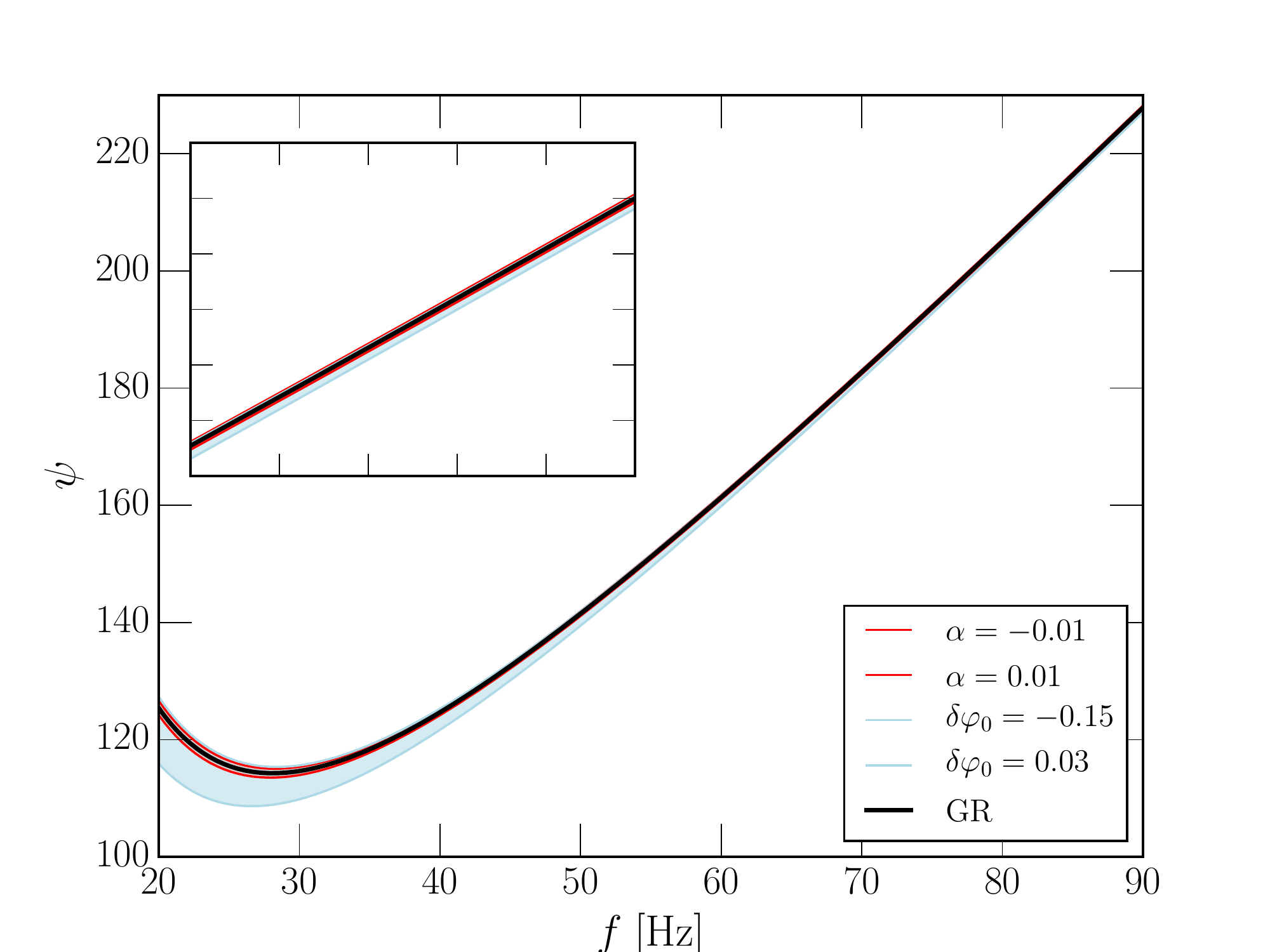}
\includegraphics[scale=0.38]{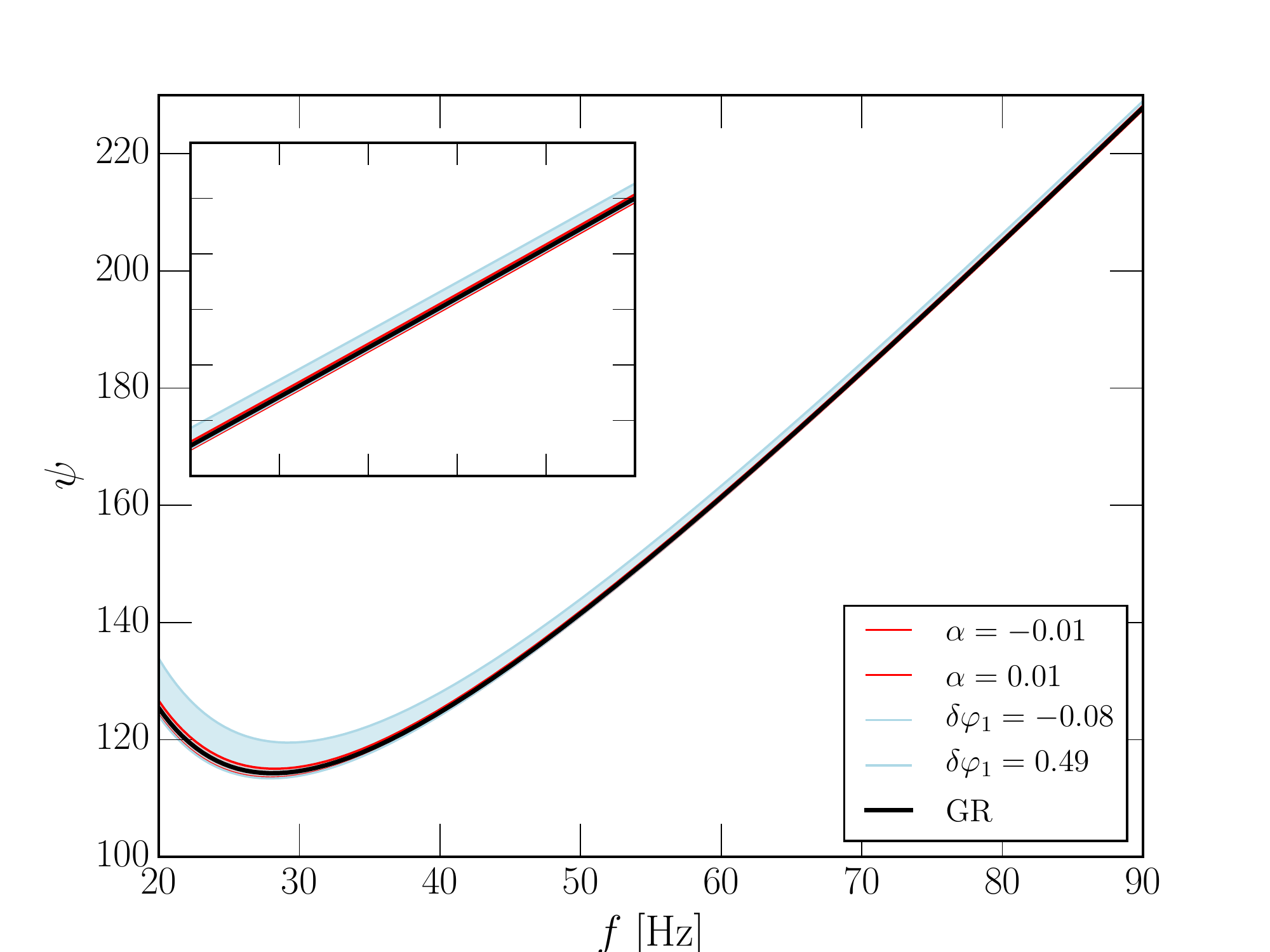}
\includegraphics[scale=0.38]{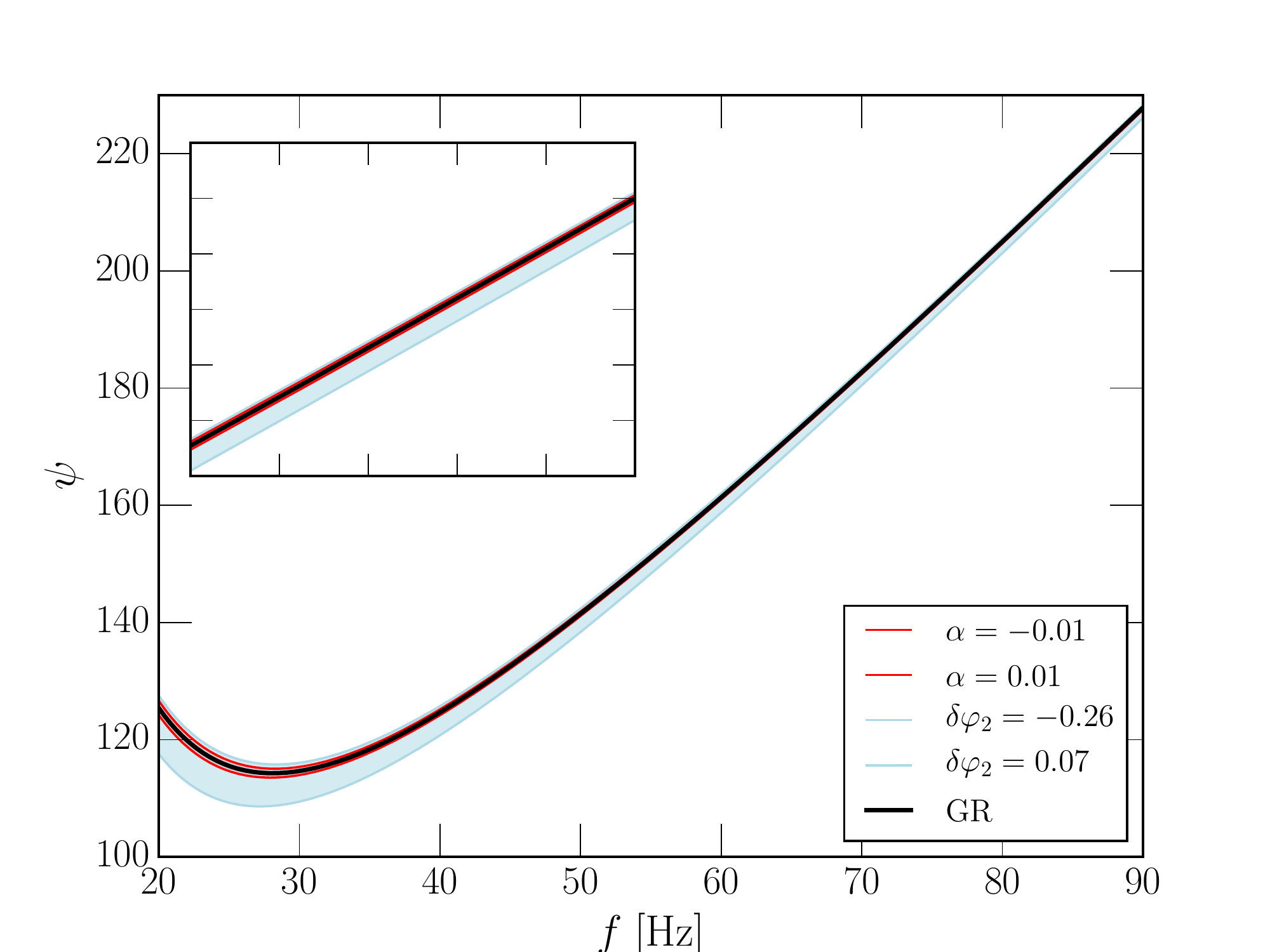}
\includegraphics[scale=0.38]{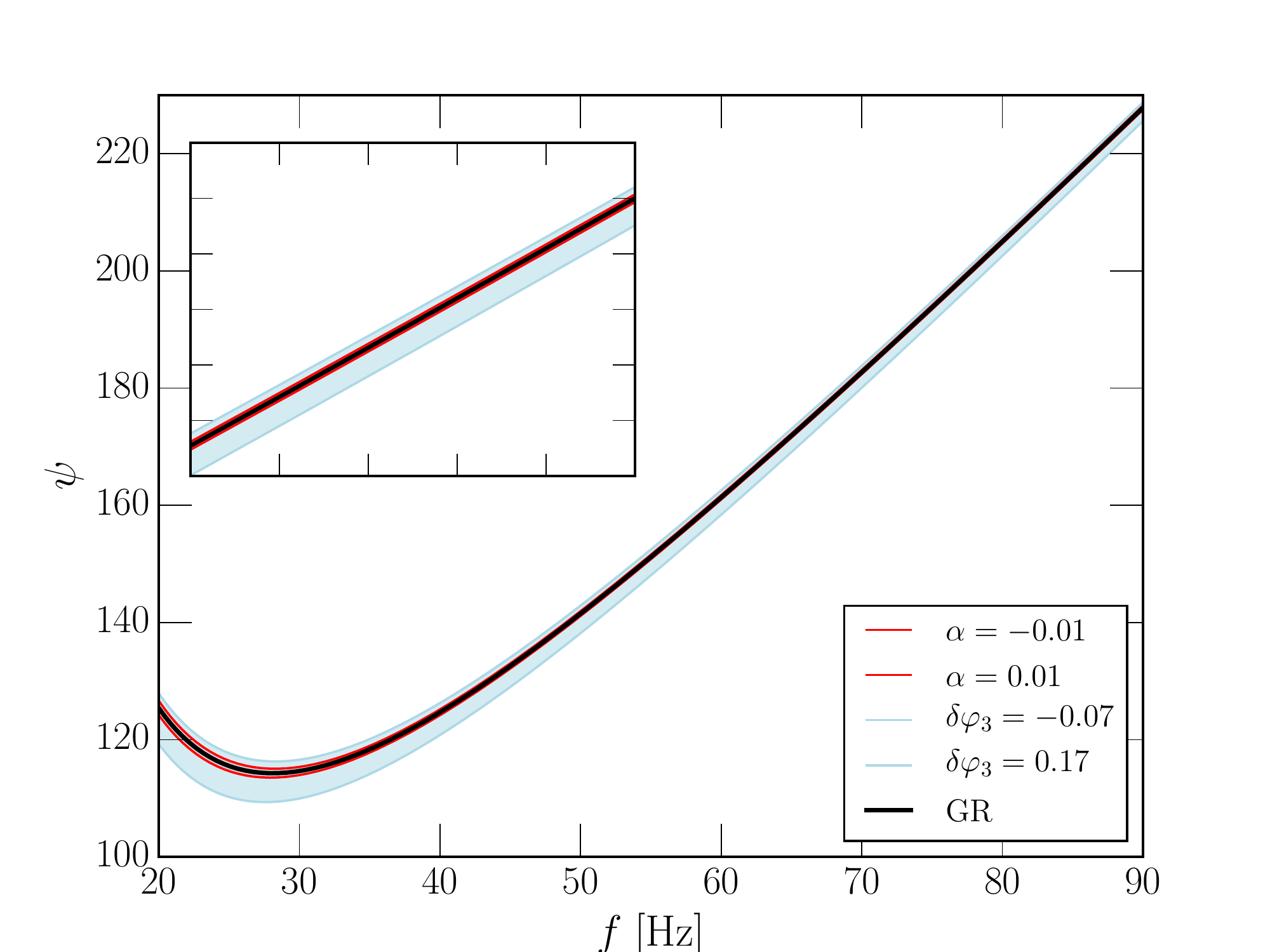}
\includegraphics[scale=0.38]{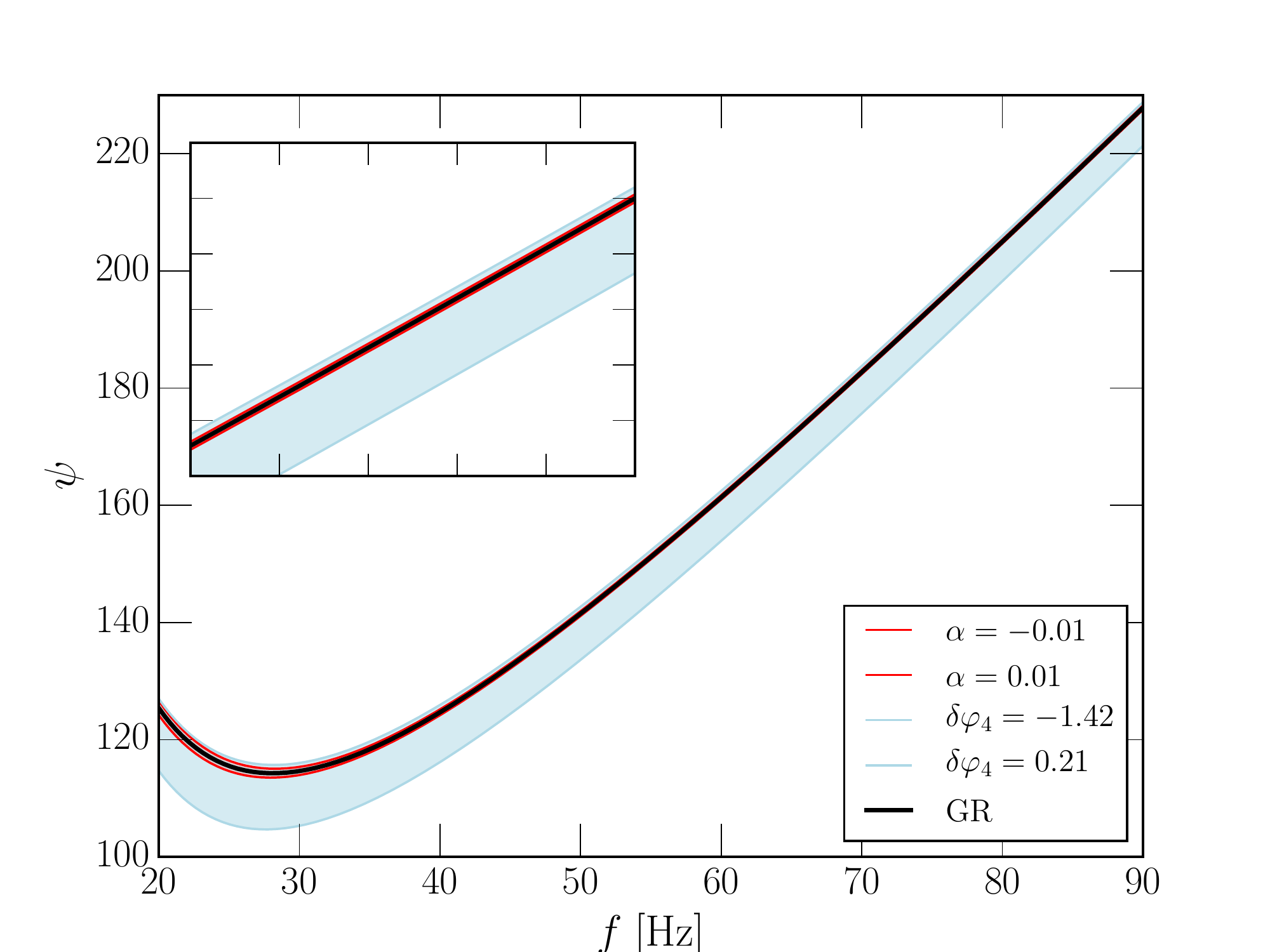}
\includegraphics[scale=0.38]{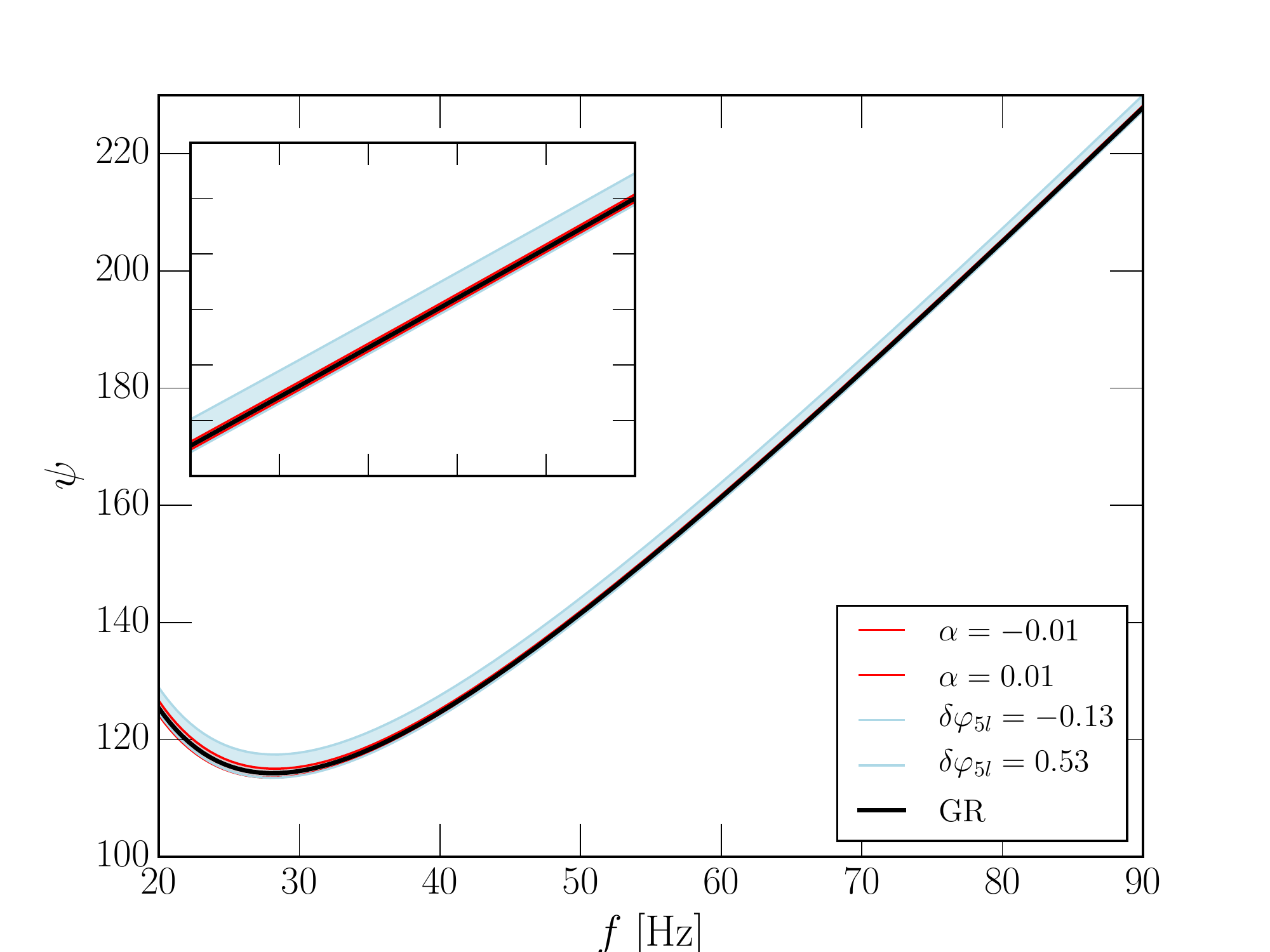}
\includegraphics[scale=0.38]{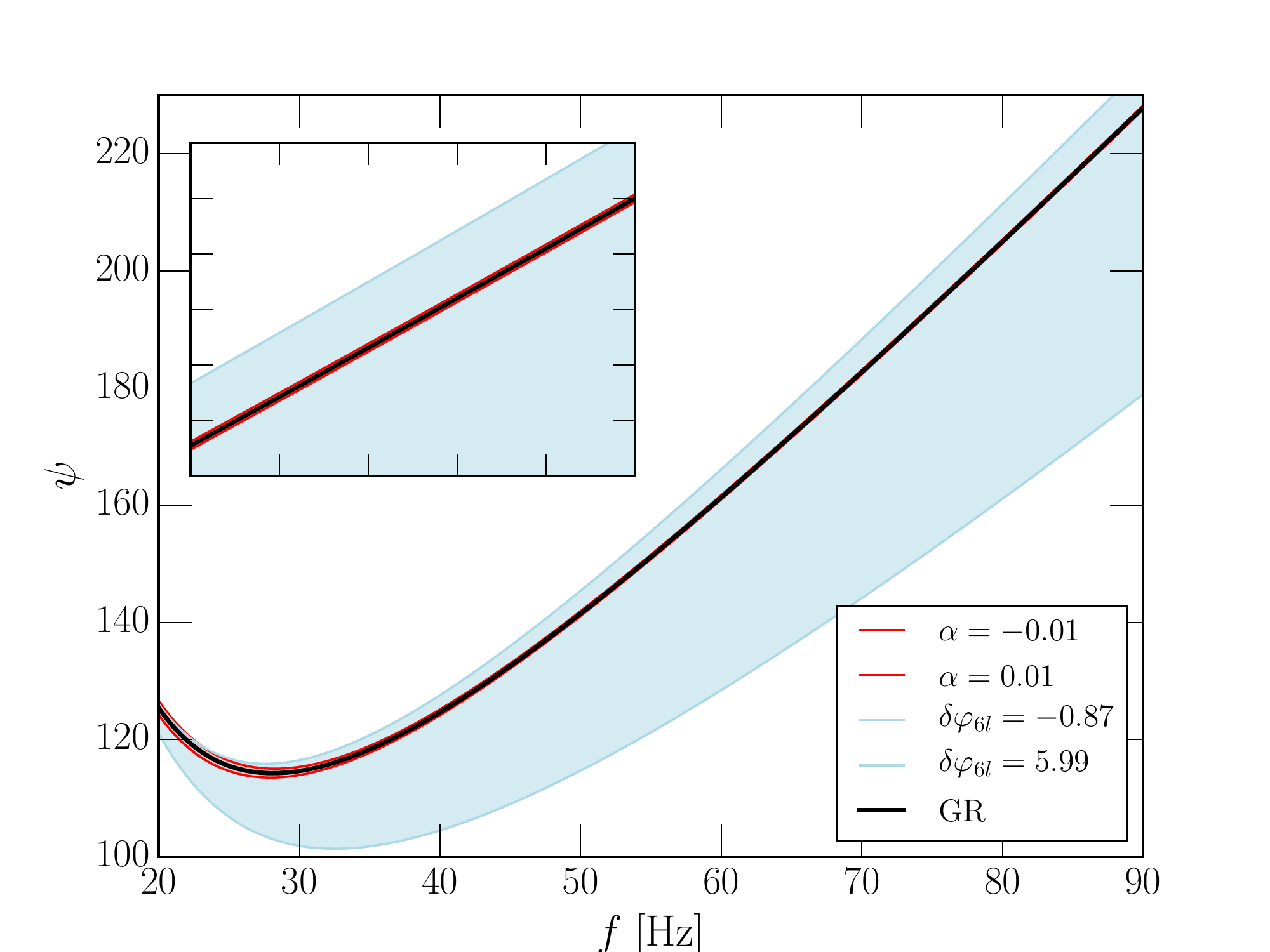}
\includegraphics[scale=0.38]{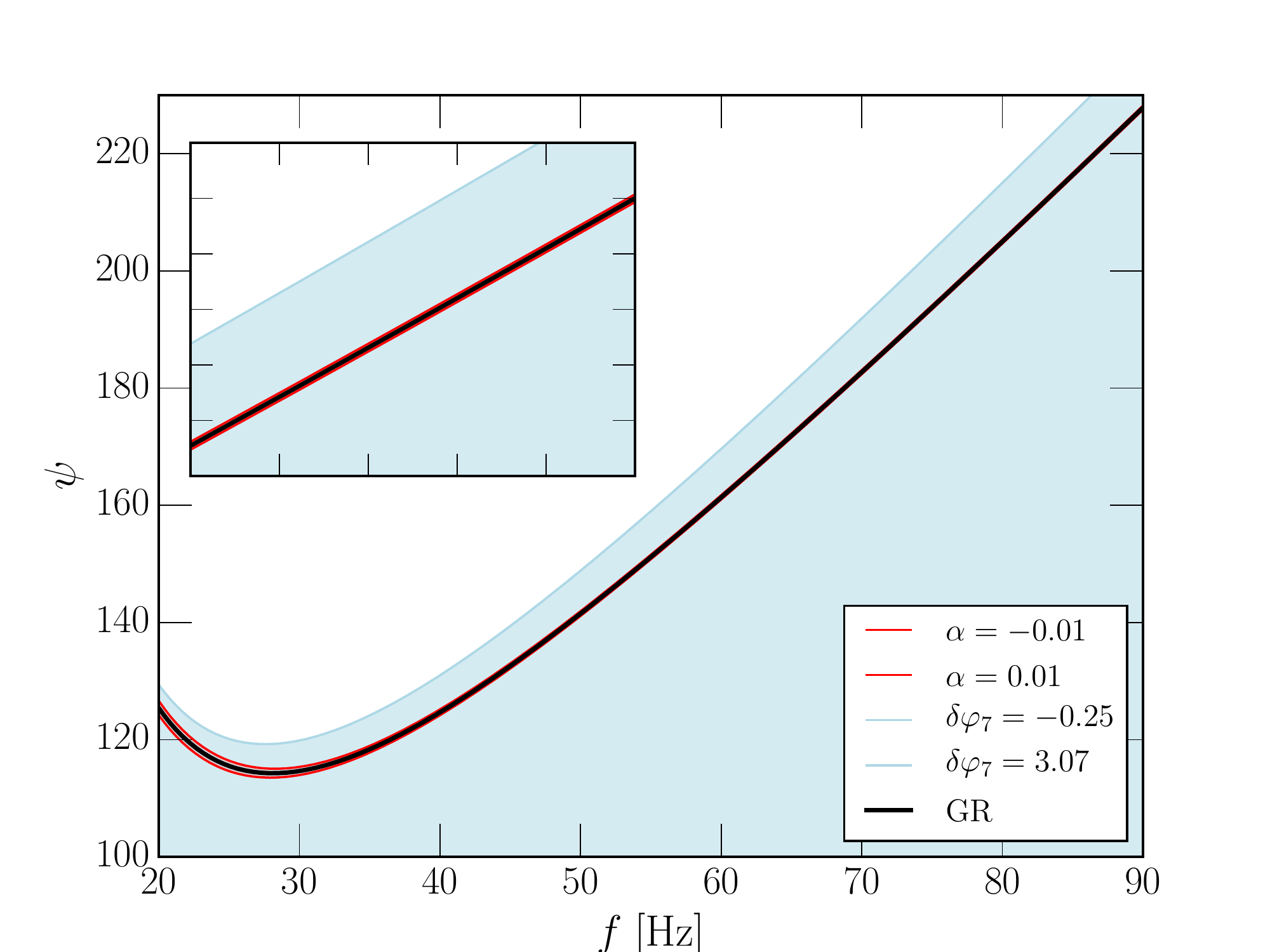}
\caption{\footnotesize{Frequency-domain phase representation for GW$150914$ with masses $m_{1} = 36.2 M_{\odot}, m_{2} = 29.1 M_{\odot}$ and initial spins $\chi_{1}<0.7, \chi_{2}<0.8$ \cite{LIGO16c} .
The solid black curve is GR prediction where $\alpha =0$ and $\delta\varphi_{i} =0$.
The red lines are $\alpha = \pm 0.01$.
The shaded blue area is the range allowed for $\delta\varphi_{i}$ parameter in accordance
with Table 1 of \cite{LIGO16a}. From left to right:  the first on the left
column show the phase at $0$PN order and the right one is for the $0.5$PN order.
The second on the left column is $1$PN while on the right there is the $1.5$PN.
The third on the left column represents the phase at $2$PN.  On the right column, the
phase at $2.5$PN is shown. Finally, the fourth on left column show the $3$PN and
on the right $3.5$PN. Note that the error on the $\delta\varphi_{7}$ is so large
that it falls outside the scale. The inset frequency ranges from $80$ Hz to $90$
Hz to be illustrate the curves at these frequencies.}}

  \label{fig:1}
\end{figure*}

\begin{figure*}
\includegraphics[scale=0.38]{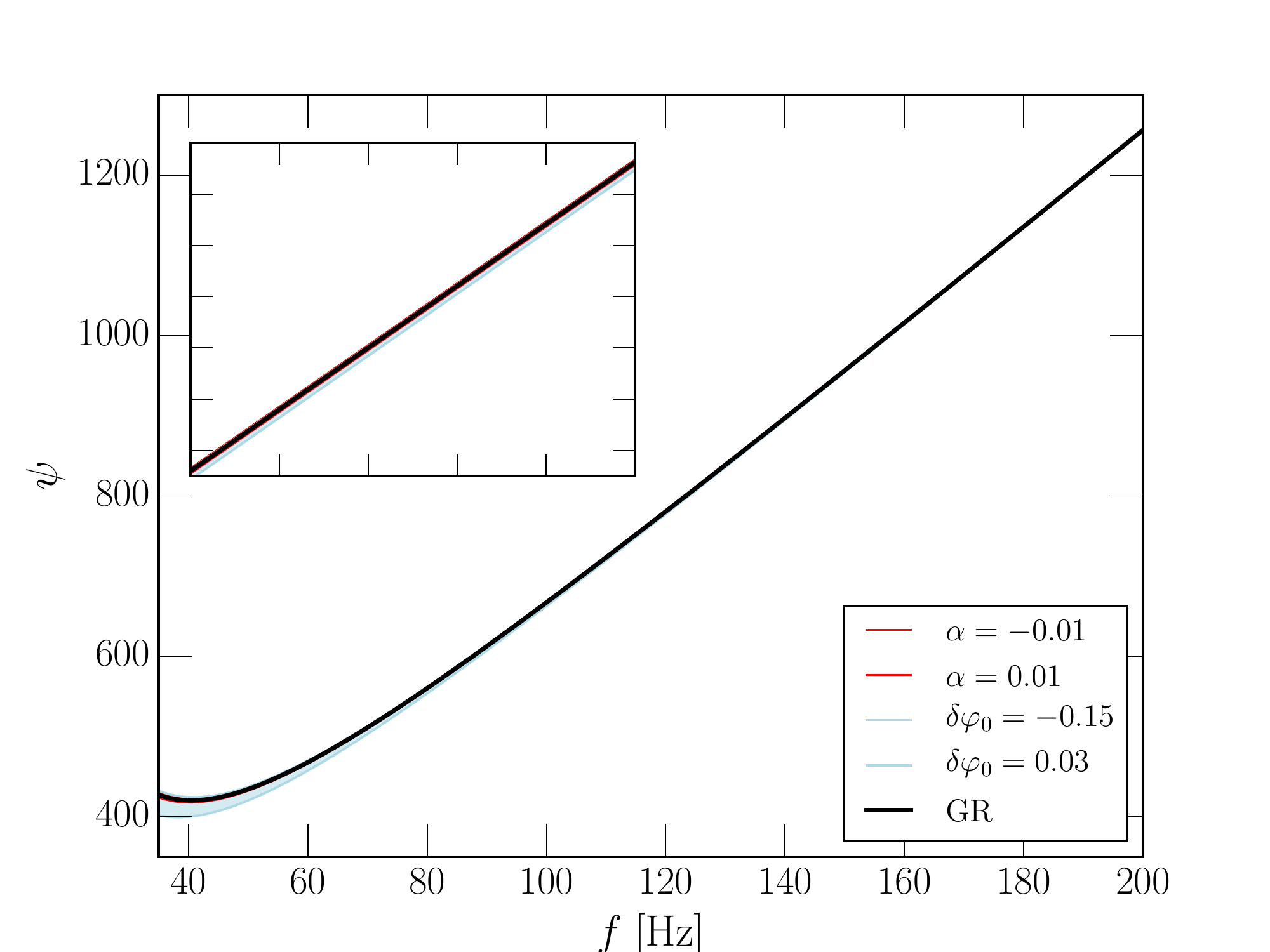}
\includegraphics[scale=0.38]{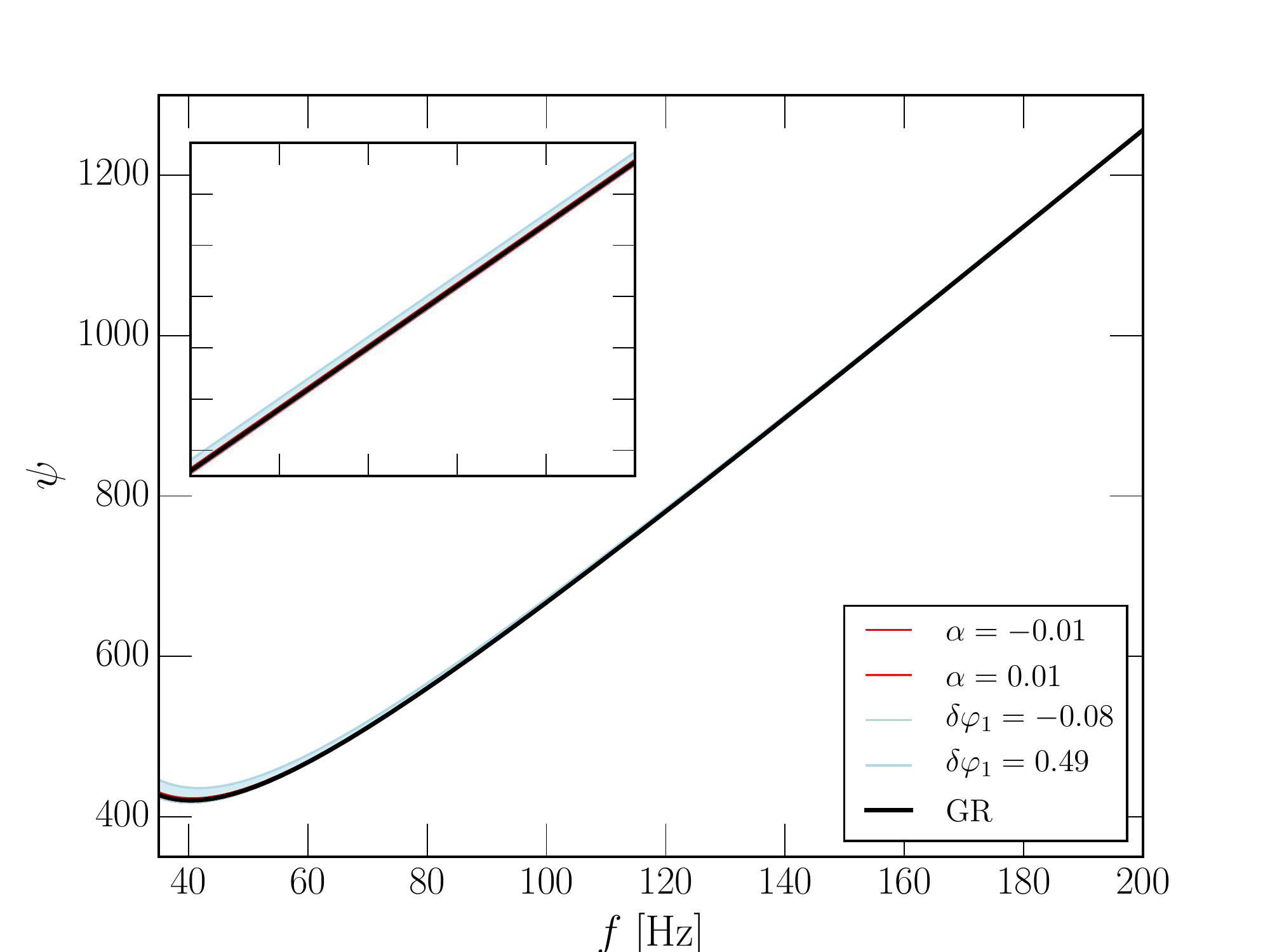}
\includegraphics[scale=0.38]{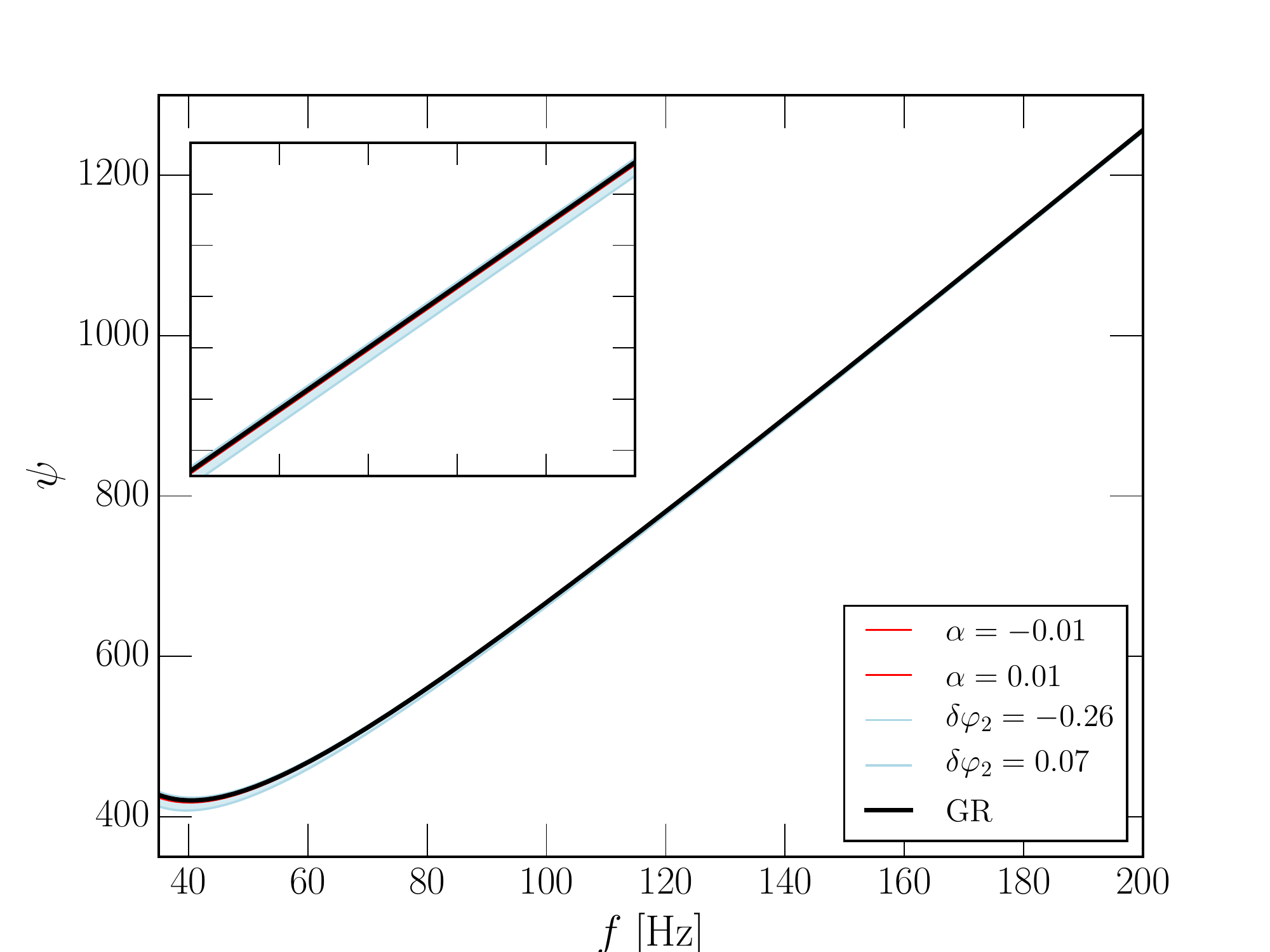}
\includegraphics[scale=0.38]{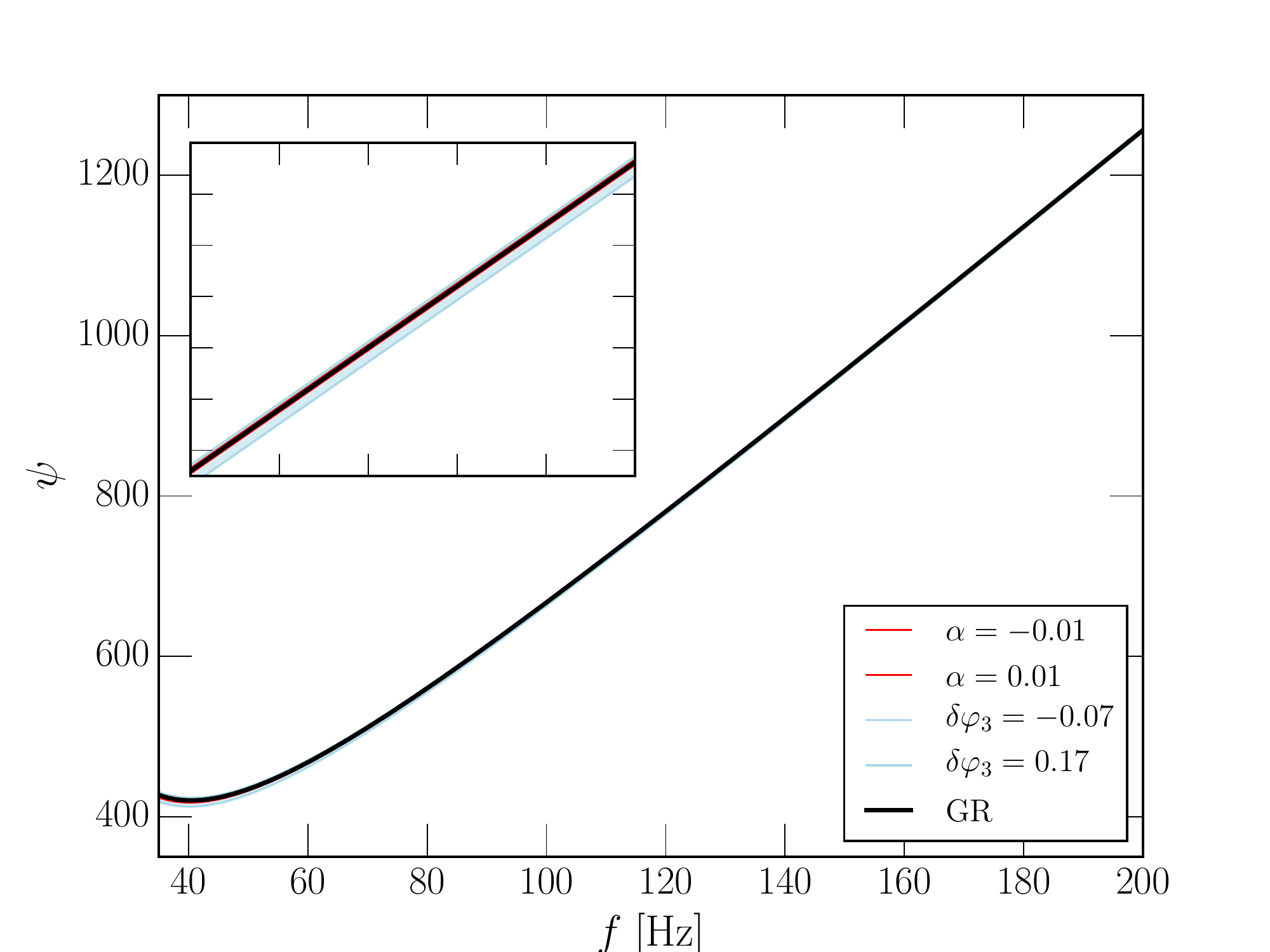}
\includegraphics[scale=0.38]{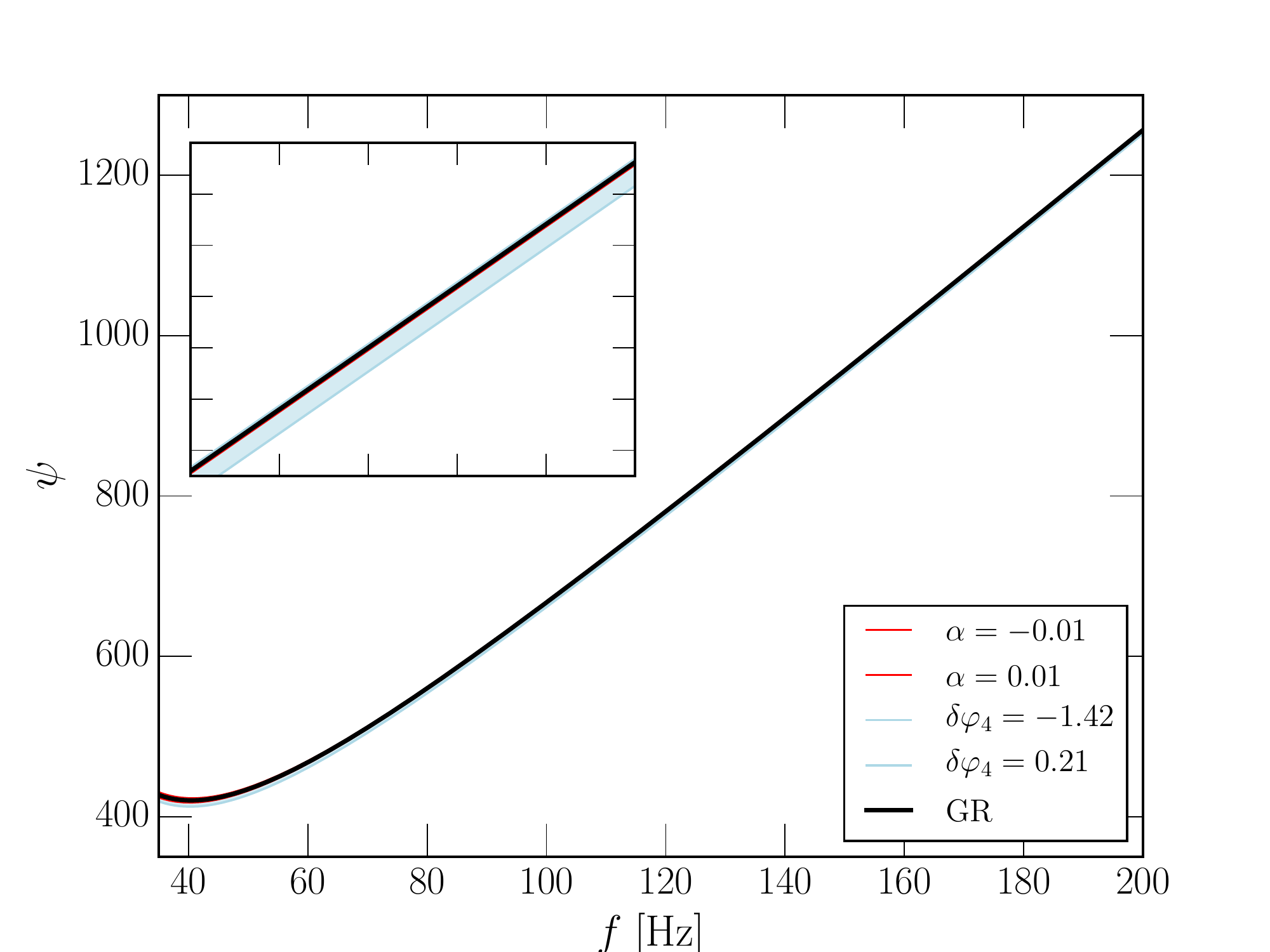}
\includegraphics[scale=0.38]{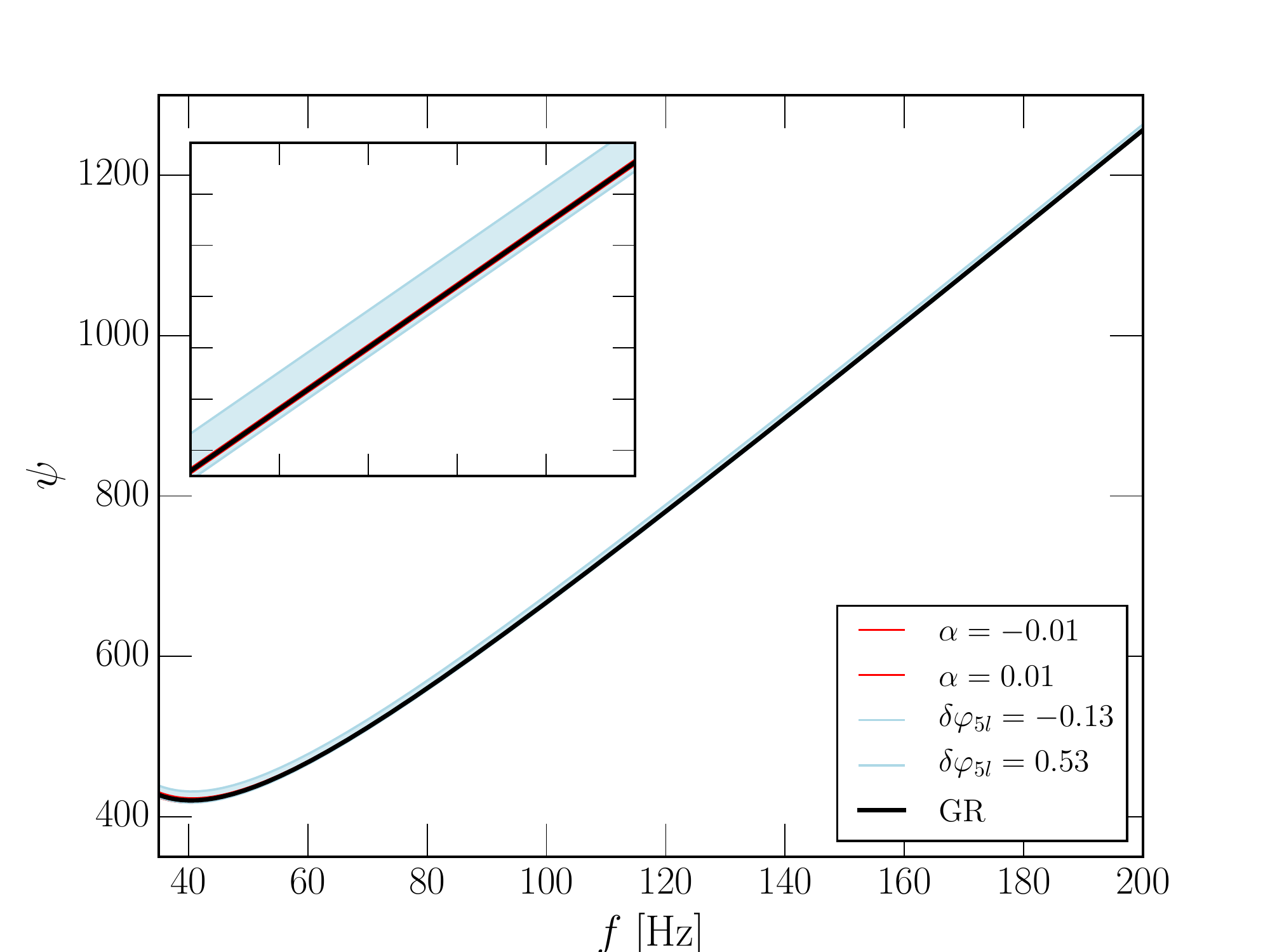}
\includegraphics[scale=0.38]{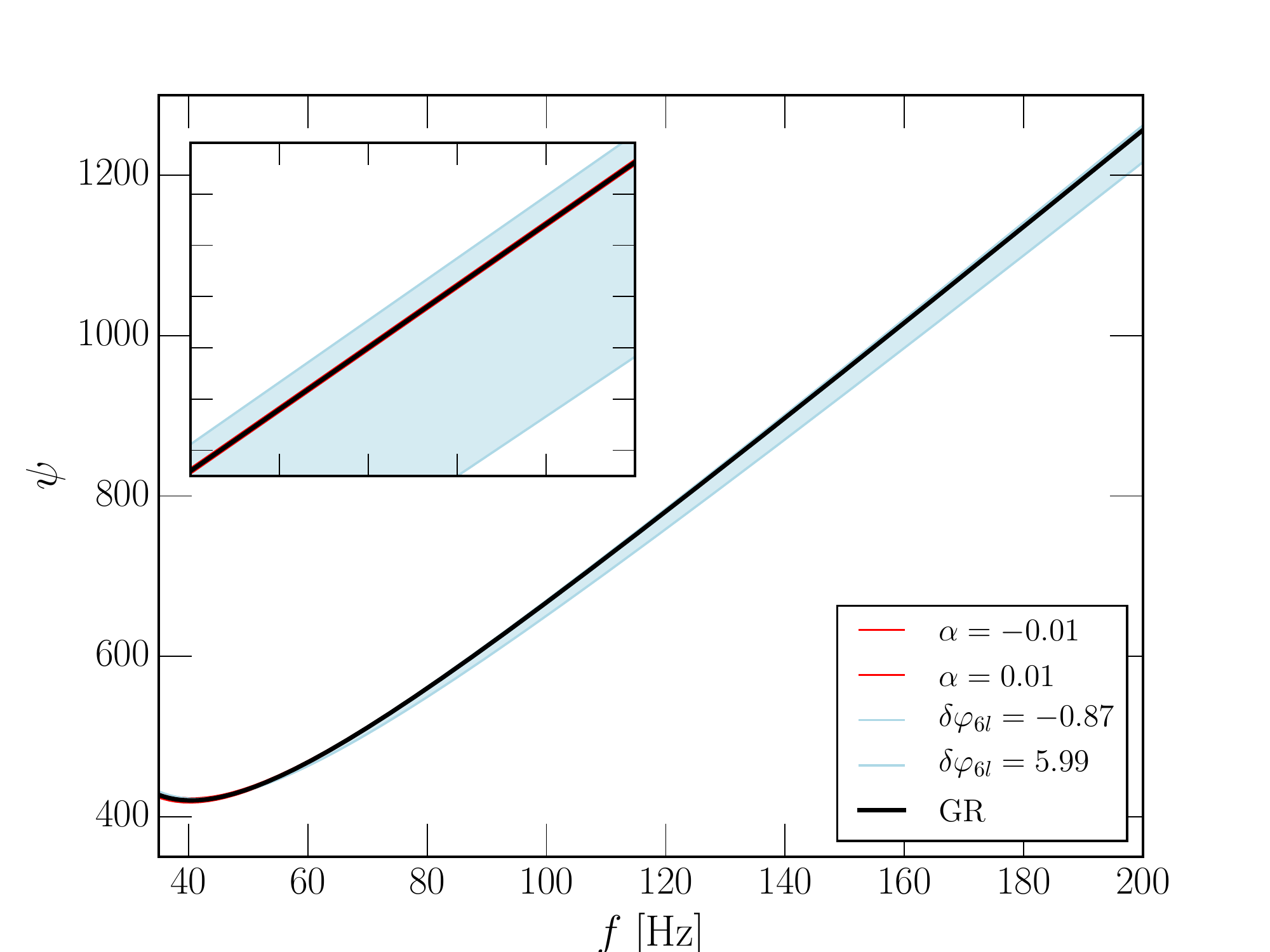}
\includegraphics[scale=0.38]{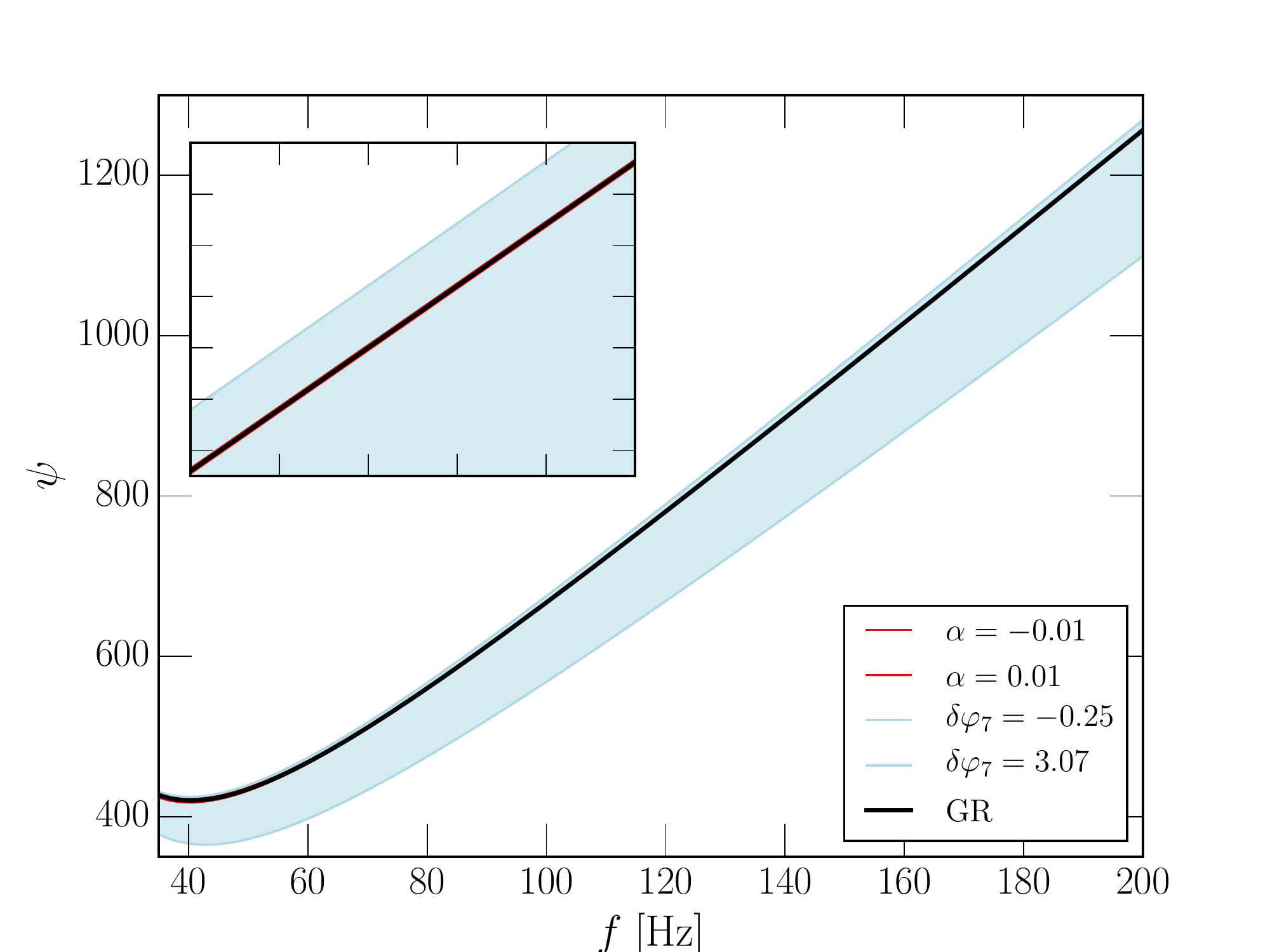}
\caption{\footnotesize{Frequency-domain phase representation for GW$151226$ with masses $m_{1} =14.2M_{\odot}\,,m_{2} =7.5 M_{\odot}$ and initial spins $\chi_{1}<0.7, \chi_{2}<0.8$ \cite{LIGO16e,LIGO16f} .
The solid black curve is GR prediction where $\alpha =0$ and $\delta\varphi_{i} =0$.
The red lines are $\alpha = \pm 0.01$.
The shaded blue area is the range allowed for $\delta\varphi_{i}$ parameter in accordance
with Table 1 of \cite{LIGO16a}. From left to right:  the first on the left
column show the phase at $0$PN order and the right one is for the $0.5$PN order.
The second on the left column is $1$PN while on the right there is the $1.5$PN.
The third on the left column represents the phase at $2$PN.  On the right column, the
phase at $2.5$PN is shown. Finally, the fourth on left column show the $3$PN and
on the right $3.5$PN. The inset shows the frequency from $180$ to $190$ Hz.
  }}
  \label{fig:2}
\end{figure*}

In Figs. \ref{fig:1} and \ref{fig:2} we have plotted the frequency domain phase representation of GW$150914$ and GW$151226$ and show the effect of varying the $\delta\varphi_{i}$ parameters as provided by the single parameter analysis of \cite{LIGO16a,LIGO16e,LIGO16f}.  Note that we follow their naming convention in introducing the quantities $\varphi_{5l}$ and $\varphi_{6l}$ that contain the logarithmic dependence with frequency.  
In addition, the variation with the leading order deviation $\alpha=\pm 10^{-2}$ is shown. The single parameter
analysis of \cite{LIGO16e} was performed by setting all but the considered
$\delta\varphi_{i}$ to 0. In contrast, the multiple parameter analysis was done
by allowing all $\delta \varphi_{i}$ to vary freely. 
The latter leads to an error that is almost one order of magnitude larger due to the additional degrees of freedom. 
We do not consider GW$150914$ and GW$151226$ data for the multiple parameter analysis performed in
\cite{LIGO16a,LIGO16e} due to the large error bars in the $\delta\varphi_{i}$ parameters.

For the masses, we have used the values as given in \cite{LIGO16f} with $m_{1} = 36.2 M_{\odot}, m_{2} = 29.1 M_{\odot}$ for GW$150914$, while $m_{1} =14.2M_{\odot}\,,m_{2} =7.5 M_{\odot}$ for GW$151226$ .
The initial spins were only constrained to be less than $\chi_{1}<0.7,
\chi_{2}<0.8$ \cite{LIGO16c,LIGO16f} and for our analysis we have taken the
values of $\chi_{1}=0.7,\chi_{2}=0.8$.  Additionally, we adopted the values $t_{c}=0.43\textrm{s}$ for GW$150914$ and $t_{c}=1\textrm{s}$
for GW$151226$ \cite{LIGO16a,LIGO16b,LIGO16e} and set $\varphi_{c}=0$ for
both.

 Furthermore, we studied the sensitivity of our results by varying the initial
masses $m_{1},m_{2}$ and initial spins $\chi_{1},\chi_{2}$ within the errors and
ranges obtained by~\cite{LIGO16c}. We found that the resulting changes were mostly
quantitative, such as altering the slopes of the curves, and that the
qualitative behaviours of the curves, such as the width of the constraints given by different
$\alpha$, remained unchanged. Thus the results plotted in Figs. \ref{fig:1} and \ref{fig:2} are
representative of the possible physical parameters reported in \cite{LIGO16c}.

In Figure ~\ref{fig:1}, we show the frequency-domain phase representation for GW$150914$ and as shaded blue area the constraints due to the $\delta \varphi_i$ of the combined events, as provided by \cite{LIGO16f} and given in Table \ref{tab:tiger-parameters}.  The GR evolution is shown as black solid line, while the extended theory is marked in the range of $\alpha\in[-10^{-2},+10^{-2}]$ as red curves.  
We report on the early inspiral range $f\in[20,90]\rm Hz$ and zoom in on the range $f\in[80,90]\rm Hz$ in the inset.  
As on can observe, at all the parameter orders the single parameter analysis does not rule out $|\alpha|<10^{-2}$.

The data of the second event, GW$151226$ is shown in Figure~\ref{fig:2}, due to the lower masse involved in the merger, the frequency is much higher.  Consequently, the inspiral regime lasts until $450 \rm Hz$ and we show the phase in the range  $f\in[40,200]\rm Hz$.  Inspecting equations \eqref{eq:phi5} and \eqref{eq:phi6}, we see that the variation with $\alpha$ is more pronounced for objects with lower total mass $M$.  Hence, in principle this event could yield tighter constraints on the value of the allowed $\alpha$.

\begin{figure*}
\includegraphics[scale=0.42]{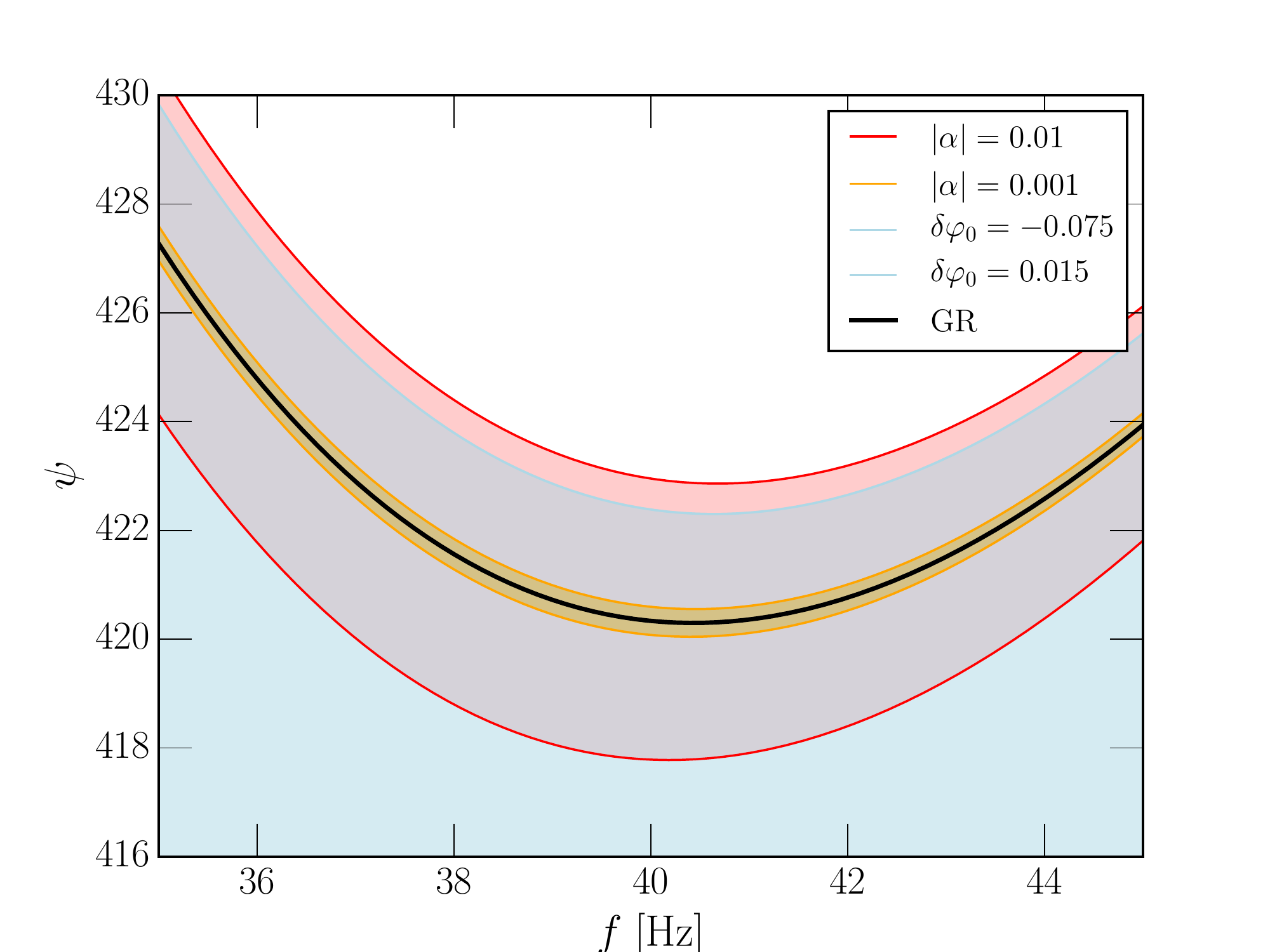}
\includegraphics[scale=0.42]{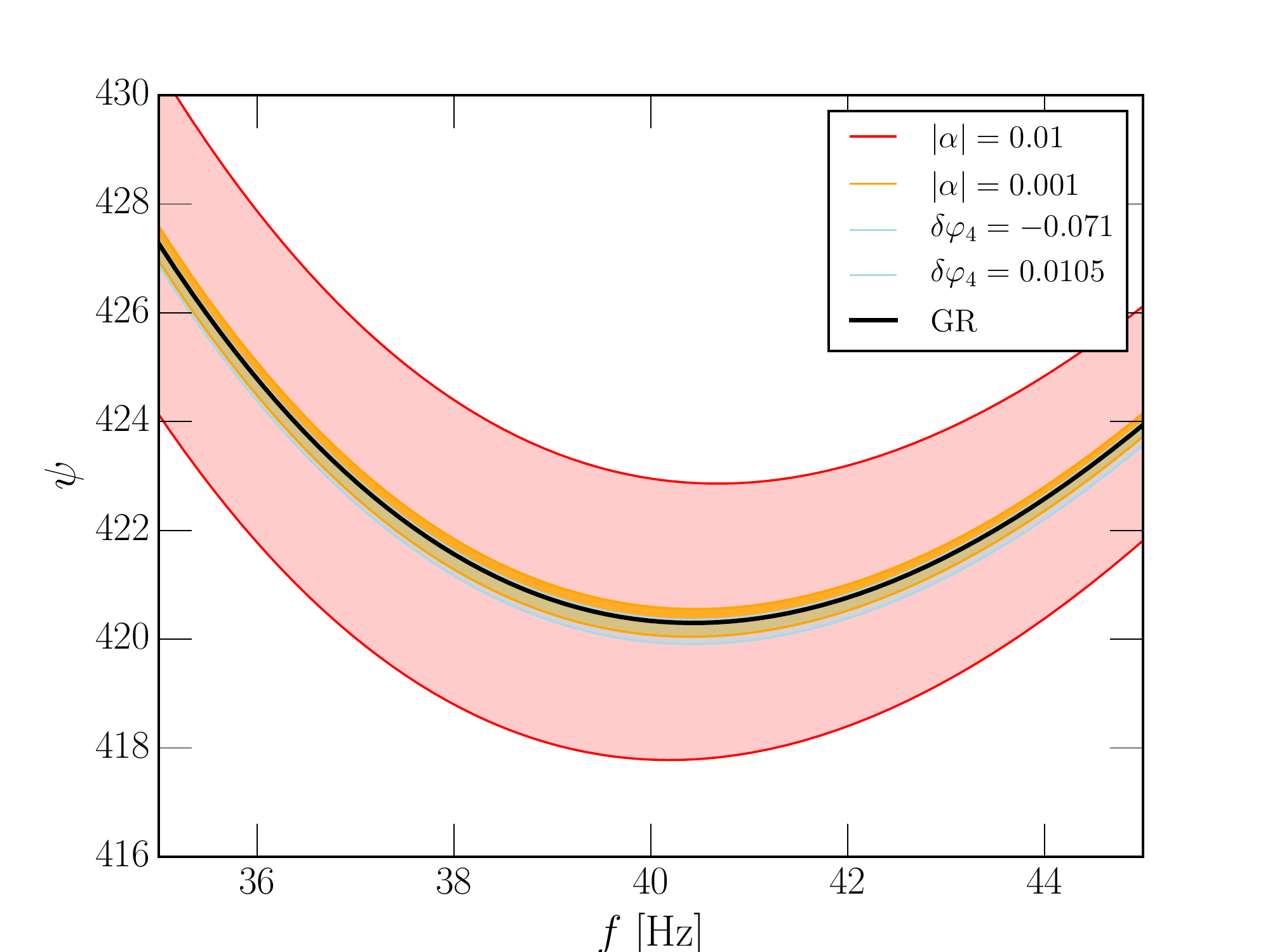}
\includegraphics[scale=0.42]{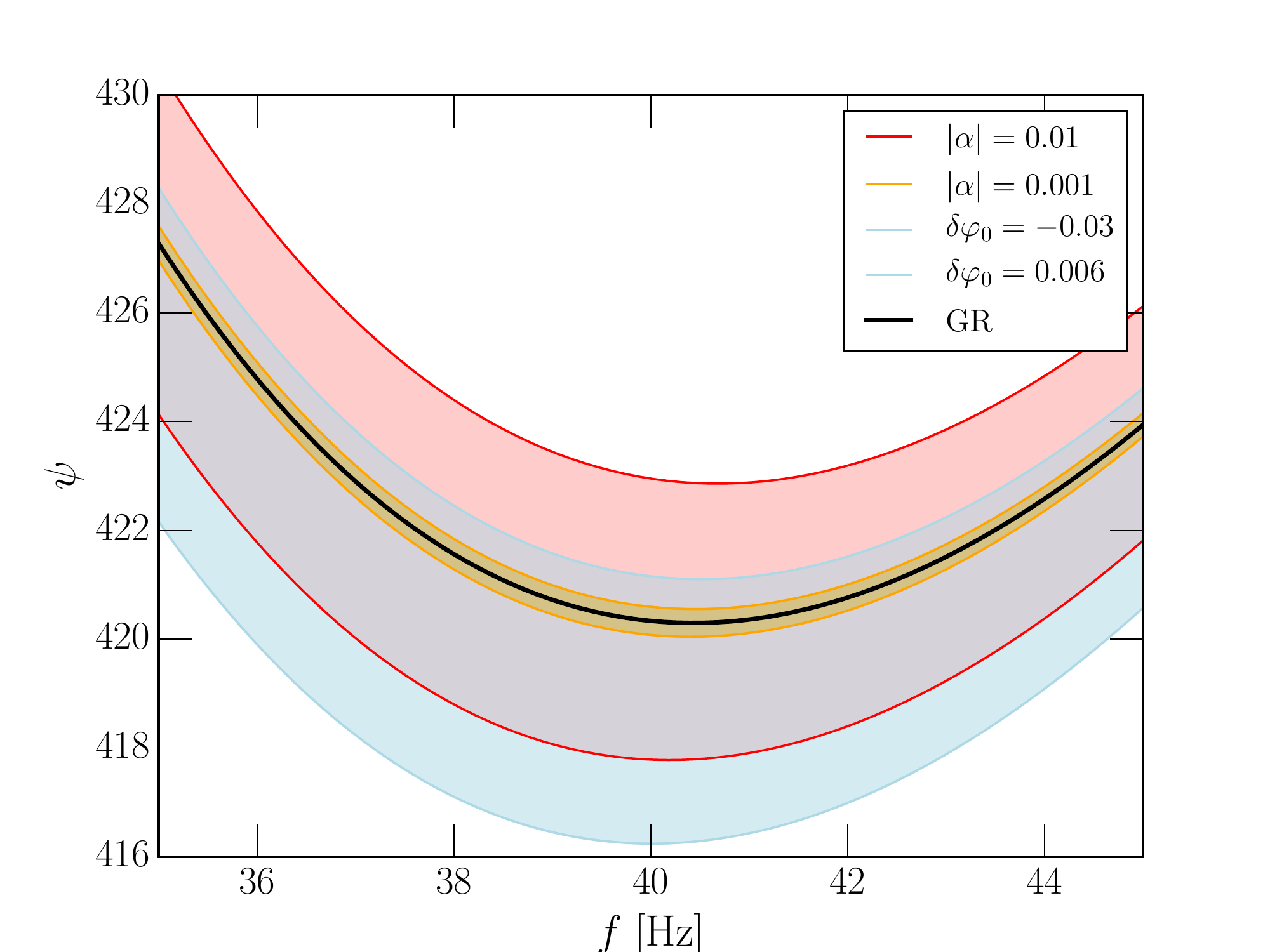}
\includegraphics[scale=0.42]{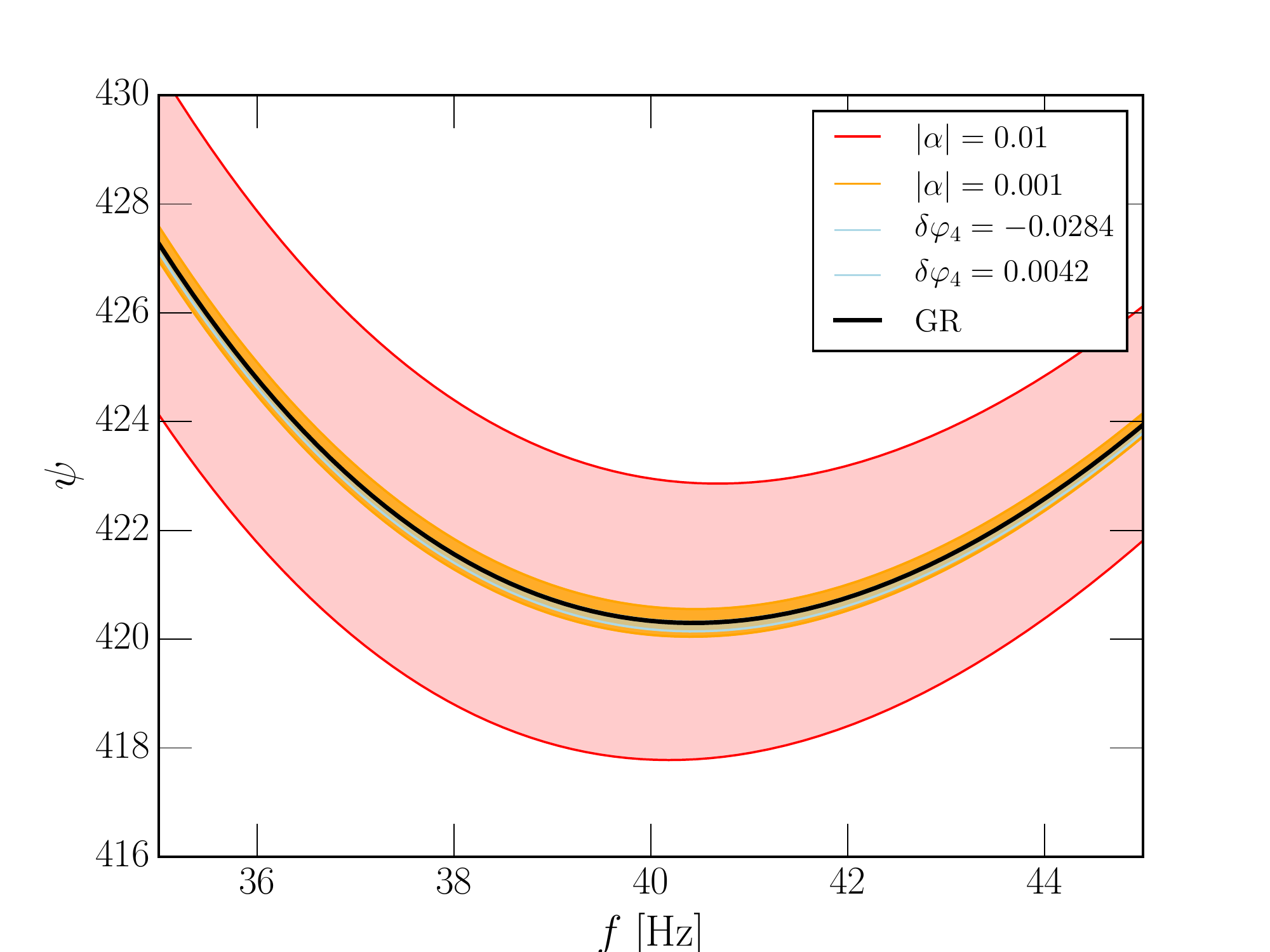}
\includegraphics[scale=0.42]{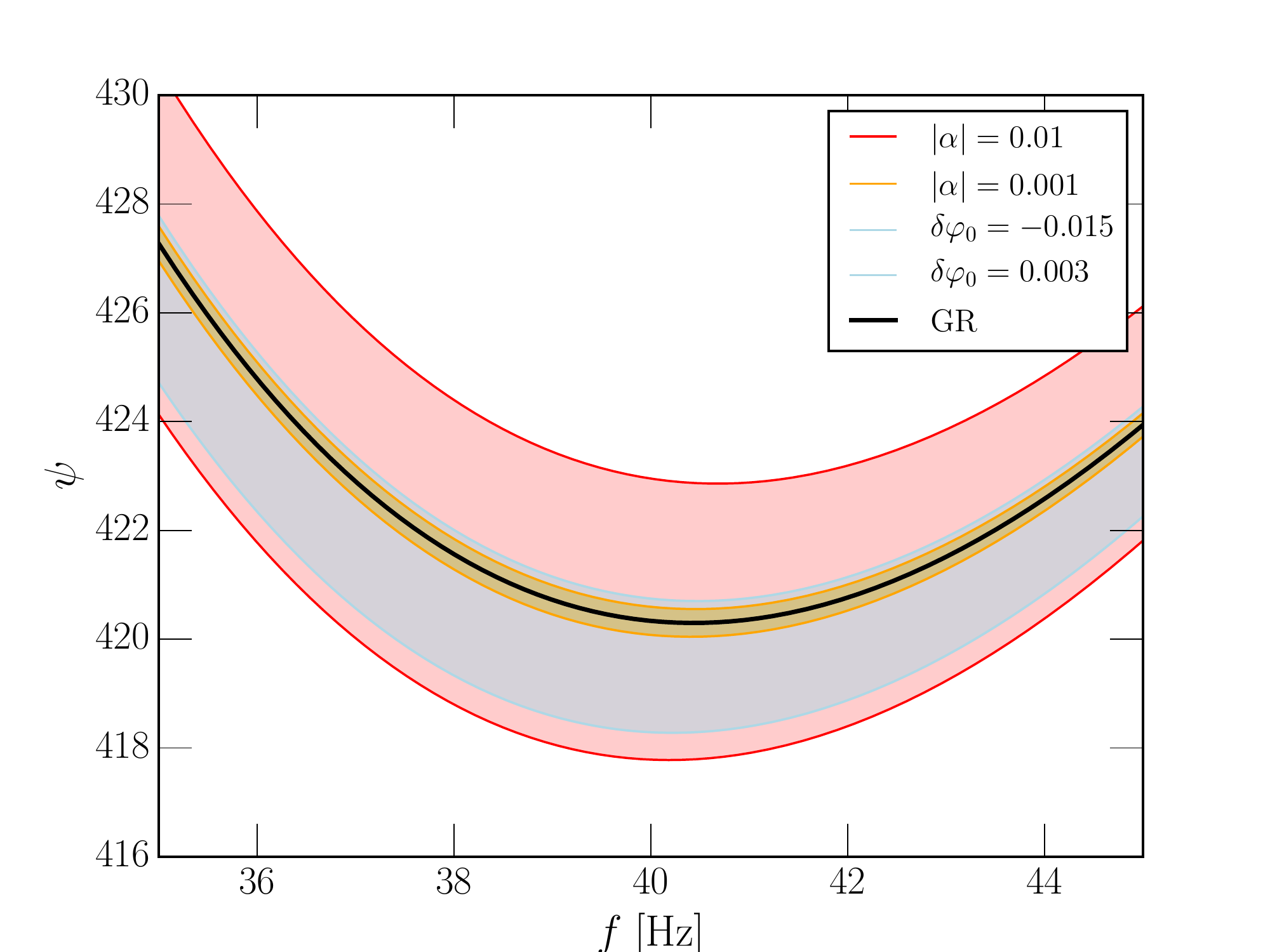}
\includegraphics[scale=0.42]{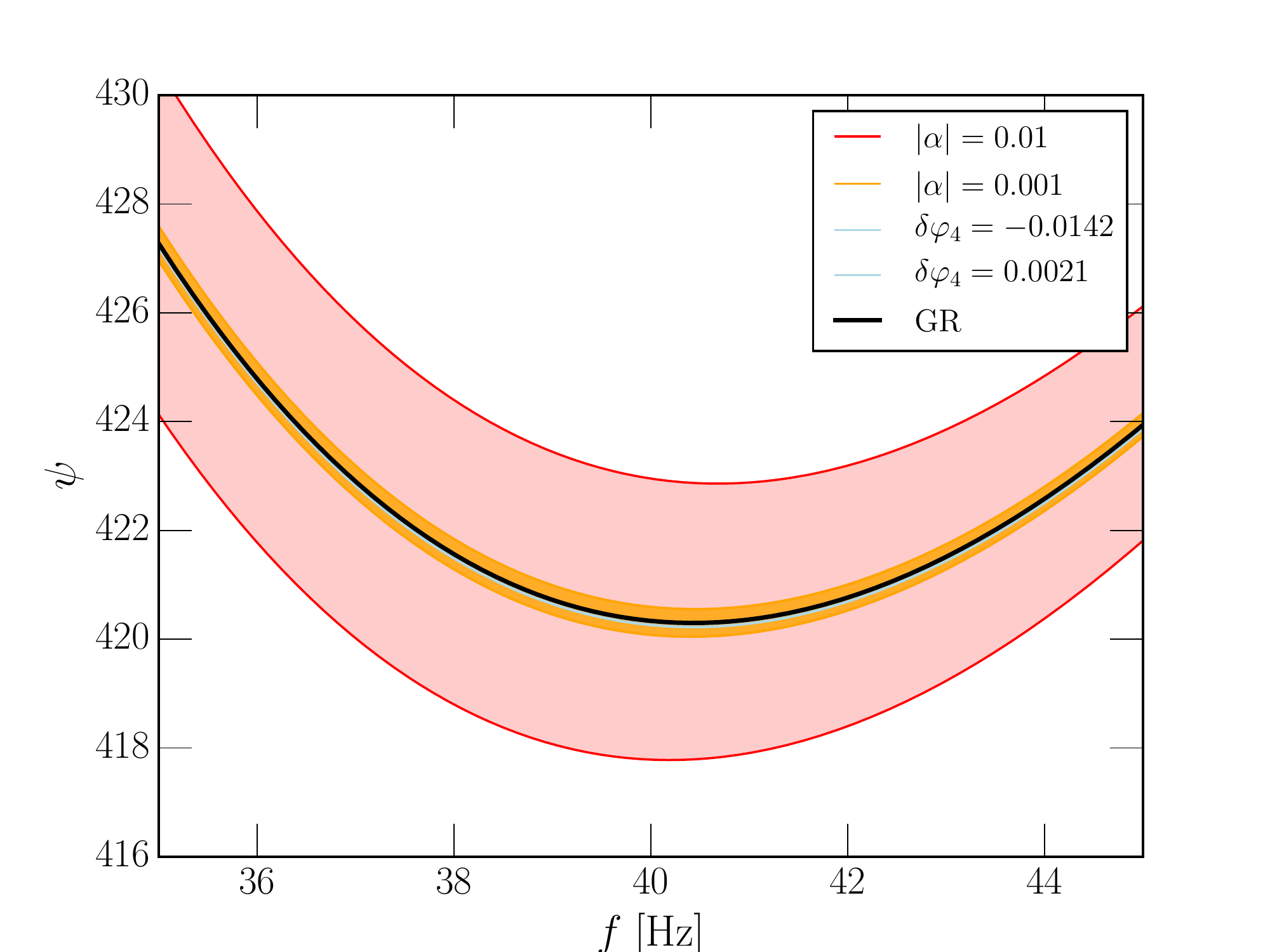}
\caption{\footnotesize{Two PN orders for GW$151226$, $0$ (left) and $4$ (right), with errors in
$\delta\varphi_{i}$ improved (from top to bottom) by factors of $2,5$, and $10$. The scales have increased in order to more clearly show the
curves.}}
  \label{fig:3}
\end{figure*}

In order to investigate the required tolerances, in Fig.~\ref{fig:3} we plot two PN orders, $0$ and
$4$ where, for GW$151226$, we have decreased the variations by factors of $2,5$ and $10$. For the PN terms of order $0$, an increase by an order of magnitude
would be sufficient to set constraints on $\alpha$. More promising still are the
higher order terms which, even with a factor of $5$ improvement, would be able to
set constraints on $|\alpha|< 10^{-3}$.

We stress that these results are obtained from the single parameter analysis, while a more correct treatment would have to adopt the uncertainties of the multiple parameter analysis.  
Furthermore, we expect that as more GWs are detected, the statistics on the combined posterior density distributions \cite{LIGO16f} will improve and we will be able
to set stronger constraints on these types of alternative theories of gravity in the
near future.  

\begin{table*}[t]
\centering
\caption{We report the frequency dependence of each parameter of Figure 6. in \cite{LIGO16f}, median and 90\% credible
  regions.  For each parameter we report the corresponding
quantities for the combined signals of GW$150914$ and GW$151226$ analyses as in \cite{LIGO16a,LIGO16f}.}
\label{tab:tiger-parameters}
\begin{tabular}{l|cccccccc}
\hline\hline
\multicolumn{1}{c|}{waveform regime} & & &
\multicolumn{1}{c}{median}\\
&parameter&$f-$dependence&GW150914+GW151226\\
\hline\hline
\multirow{9}{*}{early-inspiral regime}&
$\delta{\varphi}_0$  & $f^{-5/3}$ &  \dchizeromedian   \\
&$\delta{\varphi}_1$  & $f^{-4/3}$ &  \dchionemedian \\%
&$\delta{\varphi}_2$  & $f^{-1}$ &  \dchitwomedian \\%
&$\delta{\varphi}_3$  & $f^{-2/3}$ &  \dchithreemedian \\
&$\delta{\varphi}_4$  & $f^{-1/3}$ &  \dchifourmedian \\
&$\delta{\varphi}_{5l}$  & $\log(f)$ & \dchifivelmedian\\
&$\delta{\varphi}_6$  & $f^{1/3}$ &  \dchisixmedian   \\
&$\delta{\varphi}_{6l}$  & $f^{1/3}\log(f)$ &   \dchisixlmedian  \\
&$\delta{\varphi}_7$  & $f^{2/3}$ &\dchisevenmedian\\
\hline
\hline
\end{tabular}
\end{table*}

\section{Constraints from the  Shapiro delay}\label{quattro}

In this section we obtain constraints from the Shapiro delay using the relative time difference between observations at multiple frequencies. This allows one to infer violations of the EP  using the observed time delay from astrophysical particle messengers like photons, gravitons or neutrinos \cite{Wu16,Kahya16,Wei15a}.  To date, the strongest constraints on the frequency dependence of the PPN-$\gamma$ parameter are obtained by observations of fast radio bursts (FRB's) yielding $\Delta \gamma(f) \sim 10^{-9}$. In the case of FRB's, the largest uncertainty is the signal dispersion due to the poorly understood line-of-sight free electron population \cite{Wei15a}.  The fact that this uncertainty is completely avoided for gravitational waves makes them an appealing messenger to test EP violations.

The Shapiro gravitational time delay is caused by the slowing passage of light as it moves through a gravitational potential \eqref{eq:yukawa1} as
\begin{equation}
\Delta t_{grav}=-\frac{1+\gamma}{c^3}\int^{r_o}_{r_e}\Phi(r)dr\, , \label{eq:shapiro}
\end{equation}
where $\gamma$ is the (theory dependent) PPN parameter and $r_o$ and $r_e$ are the positions of the observer and the source of emission.
Let us conservatively assume a short burst of emission, that is all wave-frequencies are emitted at the same instant.  Now given the observed signal duration for GW150914 of $\sim 0.2 s$, we can obtain an estimate for the frequency dependence of $\gamma$, respectively $\alpha$.  In absence of other dispersive propagation effects, e.g. due to Lorentz invariance violation (see also \cite{GaoWu15,Mattingly05}), we obtain an upper limit for $\Delta \alpha/\Delta f$.

For example, in scalar tensor theories the $\gamma$ PPN parameter is expressed in terms of non minimal coupling function of a scalar field, equivalently, in terms of the $\alpha$ parameter, that is (for more details see \cite{Capozziello05}):
\begin{equation}\label{gamma}
\gamma-1\,=\,-\frac{(f'(\varphi))^2}{f(\varphi)+2[f'(\varphi)]^2}\,=\,-2\frac{\alpha^2}{1+\alpha^2}\,,
\end{equation}

In this case, the delay (\ref{eq:shapiro}) takes the form
\begin{equation}
  \Delta t_{grav}=-(2-\frac{2\alpha^2}{1+\alpha^2})/c^3\int^{r_o}_{r_e}\Phi_{\rm N}(r) (1+\alpha e^{-r/r_1}) dr\,,
\end{equation}
of which the most important contribution comes from the term linear in $\alpha$:
\begin{equation}
  \Delta t_{grav} \simeq -2\alpha/c^3 \int^{r_o}_{r_e}\Phi_{\rm N}(r) e^{-r/r_1} dr\,.
\end{equation}

It is evident that for $r_1\gg r_e,r_o$ the value for $\Delta\alpha/\Delta f$ is just half the constraint that can be set on $\Delta\gamma/\Delta f$ using the usual Shapiro delay.  Thus with the same assumptions for the potential encountered by the gravitons as \cite{Kahya16} (corresponding to a Shapiro delay of $1800$ days), we can set the limit $|\alpha(250 Hz)-\alpha(35 Hz)|<1.3\times 10^{-9}$.


It is important to note that this seemingly tight limit relates to the frequency dependence only, e.g. if we have $\alpha=\alpha_0 + \alpha(f)$, the constant term $\alpha_0$ is entirely unconstrained by this experiment. However, knowledge of $\Delta\alpha/\Delta f$ can be used to extrapolate measurements of the absolute value of $\alpha$ across the spectrum and thus extend their range of validity.

\section{Discussion and conclusions}
\label{cinque}

The GW$150914$ and GW$151226$ signals \cite{LIGO16b, LIGO16f} show the inspiral and merger regimes and GW$150914$ is also observed in the ringdown phase. 
Here we have analysed the inspiral data for GW$150914$ and GW$151226$, using and extended post-Newtonian approximation.
We would like to underline that corrections coming from alternative gravity to the standard relativistic equations and waveforms describing binary black hole systems are negligible up to $2.5$PN order (see {\it e.g.} \cite{Mirshekari13}).

However, since, as we have demonstrated, extended theories of gravity give rise to an effective gravitational coupling constant $G_{eff}$, the post-Newtonian dynamics of any metric formalism can be obtained straightforwardly for the lowest order deviation parameter $\alpha=const.$.  
The recently detected gravitational waveforms of GW$150914$ and GW$151226$ can thus give constraints on the theory.   

Using the fact that the gravitational wave frequencies are modulated through $G_{eff}$, we have shown that this modulation
will change the phase of the detected gravitational signal.
Our conclusions are in agreement with \cite{Miller16} who found that
that GW$150914$ and GW$151226$ do not place strong constraints on the theory of gravity, since the parameters of the merging black holes are not measured with high enough precision. However, improved statistics on the deviations $\delta\phi_i$ could remedy this shortcoming in the future.  

Moreover, we have used the Shapiro delay of GW$150914$ to set an upper limit
$|\alpha(250 Hz)-\alpha(35 Hz)|<1.3\times 10^{-9}$.
Although this result was obtained for scalar tensor theories, this applies
for all theories where the PPN-$\gamma$ is at least quadratic in $\alpha$.

 The constraints provided by GW$150914$ and GW$151226$ on GR and, in general, metric theories of gravity, are unprecedented due to the nature of the sources and the strong field regime. However they have not reached
high enough precision to definitively discriminate among concurring theories.  Furthermore, in order to extract new
physical effects, one would need a wide range of GW waveforms beyond the standard forms adopted for GR and allow for polarisations beyond the standard $\times$ and + modes \cite{Bogdanos10}.

Finally,  more stringent bounds could be obtained by combining results from multiple GW observations~\cite{Ghosh16,Li12,Agathos14,Pozzo11}.
Given the rate of coalescence of binary black holes as inferred in Ref.~\cite{LIGO16d}, we are looking forward to the upcoming joint observation surveys  from advanced  LIGO and VIRGO experiments.

\section*{Acknowledgements}
The authors thank the referee for the useful comments
and suggestions that allowed improving the paper.
We acknowledge Luciano Rezzolla for fruitful discussion and his valuable comments.
This research was supported by ERC Synergy
Grant "BlackHoleCam" Imaging the Event Horizon of Black Holes awarded by the ERC in 2013 (Grant No. 610058),
by the Volkswagen Stiftung (Grant 86 866), by the LOEWE-Program in HIC for FAIR, and by ``NewCompStar'',
COST Action MP1304. AAA and BJA are also supported in part by the project
F2-FA-F113 of the UzAS and by the ICTP through the projects OEA-NET-76,
OEA-PRJ-29. AAA and BJA thank the Institut f{\"u}r Theoretische Physik for warm
hospitality during their stay in Frankfurt. M.D.L acknowledge the COST Action CA15117 (CANTATA) and INFN Sez. di Napoli (Iniziative Specifiche QGSKY and TEONGRAV).


\bibliographystyle{apsrev4-1}  
\bibliography{add-references,gravreferences}

\end{document}